\documentclass[fleqn,usenatbib]{mnras}
\usepackage{natbib}
\usepackage[T1]{fontenc}
\usepackage{commath}
\usepackage{amsmath}	% Advanced maths commands
\usepackage{amssymb}	% Extra maths symbols
\usepackage{ae,aecompl}
\usepackage{graphicx}	% Including figure files
\usepackage[usenames,dvipsnames,svgnames,table]{xcolor}
\usepackage[table]{xcolor}
\usepackage{verbatim}
\usepackage[font = small, labelfont=bf, labelsep=period]{caption}
\usepackage{subcaption}
\usepackage{breqn}
\usepackage{float}
\usepackage{dblfloatfix}

\usepackage[utf8]{inputenc}
\usepackage{multicol}
\usepackage{bibspacing}
\usepackage{lipsum}
\usepackage{hyperref}
\usepackage{diagbox}
\usepackage[inline]{enumitem}
\usepackage{enumitem}
\usepackage[normalem]{ulem}
\setlist{
	listparindent=\parindent,
	parsep=0pt,
}
\hypersetup{ %
			colorlinks=true,%
			pdfauthor={Nomen Tuum},%
			pdftitle={},%
			citecolor=blue%
			}

\def\red#1 {\textcolor{red}{#1}\ }  
\def \green#1 {\textcolor{green}{#1}\ }  

\title[Eccentric MBHB Populations]{Signatures of Circumbinary Disk Dynamics in Multi-Messenger Population Studies of Massive Black Hole Binaries}
\author[Magdalena Siwek et al.]{
Magdalena Siwek$^{1}$\thanks{E-mail: magdalena.siwek@cfa.harvard.edu}, 
Luke Zoltan Kelley$^{2}$ and
Lars Hernquist $^{1}$
\\
$^{1}$Center for Astrophysics, Harvard University, Cambridge, MA 02138, USA \\
$^{2}$ Department of Astronomy, University of California, Berkeley, 501 Campbell Hall \#3411, Berkeley, CA 94720, USA
}
\date{Accepted XXX. Received YYY; in original form ZZZ}

\pubyear{2022}

\begin{document}
\label{firstpage}
\pagerange{\pageref{firstpage}--\pageref{lastpage}}
\maketitle

%%%%%%%%%%%%%%%%%%%%%%%%%%%%
%%%%%%%% ABSTRACT %%%%%%%%%%
\begin{abstract}
	 We investigate the effect of cutting-edge circumbinary disk (CBD) evolution models on massive black hole binary (MBHB) populations and the gravitational wave background (GWB). We show that CBD-driven evolution leaves a tell-tale signature in MBHB populations, by driving binaries towards an equilibrium eccentricity that depends on binary mass ratio. We find high orbital eccentricities ($e_{\rm b} \sim 0.5$) as MBHBs enter multi-messenger observable frequency bands. The \textbf{CBD-induced eccentricity distribution of MBHB populations in observable bands is independent of the initial eccentricity distribution} at binary formation, erasing any memory of eccentricities induced in the large-scale dynamics of merging galaxies. Our results suggest that \textbf{eccentric MBHBs are the rule rather than the exception in upcoming transient surveys}, provided that CBDs regularly form in MBHB systems. We show that the \textbf{GWB amplitude is sensitive to CBD-driven preferential accretion onto the secondary}, resulting in an increase in GWB amplitude $A_{\rm yr^{-1}}$ by over 100\%  with just 10\% Eddington accretion. As we \textbf{self consistently allow for binary hardening and softening}, we show that CBD-driven orbital expansion does not diminish the GWB amplitude, and instead increases the amplitude by a small amount. We further present detection rates and population statistics of MBHBs with $M_{\rm b} \gtrsim 10^6 \, M_{\odot}$ in LISA, showing that most binaries have equal mass ratios and can retain residual eccentricities up to $e_{\rm b} \sim 10^{-3}$ due to CBD-driven evolution. \\

\end{abstract}

\begin{keywords}
black holes, gravitational waves, accretion, accretion disks, binaries, torques, hydrodynamics, transients
\end{keywords}

\section{Introduction}
Pulsar Timing Arrays (PTAs; \citealt{Foster1990}) have found strong evidence for a Gravitational Wave Background (GWB; \citealt{NANOGravDetection2023, EPTADetection2023, PPTADetection2023, CPTADetection2023}), likely originating from the inspiral of massive black hole binaries (MBHBs; \citealt{NANOGravMBHBs2023}). Concurrently, the most comprehensive study of optical transients with the Legacy Survey of Space and Time (LSST; \citealt{LSST2019}) at the Rubin Observatory is approaching. If EM signatures of MBHBs are found, the synergy of these detection methods will enable the multi-messenger study of MBHBs using transient surveys and low-frequency GWs, and elucidate the evolution and fate of MBHBs across cosmic time. 

MBHBs form as a result of galaxy mergers. The MBHs contained in the merger remnant initially evolve through dynamical friction, followed by stellar scattering \citep[e.g.,][]{Begelman1980, Quinlan1997, Sesana2006}, which is likely efficient at hardening binaries to parsec separations \citep[e.g.,][]{Yu2002, Khan2011, Gualandris2016}. At parsec scales, the formation of a circumbinary accretion disk (CBD) is likely in the gas-rich environment of a post-merger galaxy \citep[e.g.,][]{Barnes1991,Barnes1996}. In such cases, the orbital evolution of sub-parsec MBHBs may be dominated by CBD physics until the GW-driven evolution takes over. 
Partially as a result of their significance in MBHB evolution and the multi-messenger signatures expected from accreting MBHBs, CBD-driven orbital evolution has gained significant attention in the literature. Accretion from a CBD has 3 main effects on the binary evolution;

\begin{enumerate}
\item Preferential accretion onto the secondary will increase the mass ratio ($q_{\rm b}$) of the binary (e.g., \citealt{Farris2014, Duffell2020}), even in eccentric binaries \citep{Siwek2023a}, potentially leading to significant increases in the GWB amplitude \citep{Siwek2020}.
\item The semi-major axis ($a_{\rm b}$) of the binary evolves. The binary can undergo orbital decay (``hardening"; \citealt{Haiman2009}) or orbital expansion (``softening"), depending on numerous factors, including (but likely not limited to) binary mass ratio and eccentricity \citep{Siwek2023b}, as well as accretion disk hydrodynamics.
\item Dynamical interaction between the binary and the CBD leads to a stabilizing effect on the orbital eccentricity ($e_{\rm b}$), and the emergence of an ``equilibrium eccentricity" \citep[e.g.,][]{Roedig2011, DorazioDuffell2021, Zrake2021}, which depends on binary mass ratio \citep{Siwek2023b}.
\end{enumerate}

CBD-driven orbital hardening is likely a sub-dominant mechanism in determining the merger timescales of MBHBs \citep[e.g.,][]{Kelley2017,Bortolas2021}. However, the effect of accretion and orbital eccentricity induced by the CBD could imprint itself on the population statistics of MBHBs observed in electromagnetic (EM) and gravitational wave (GW) observatories. 
Identifying signatures of CBD physics requires population synthesis modeling of MBHB populations, and assumptions about their initial mass ratio and eccentricity distributions. 

The initial mass ratio distribution of MBHBs is set by the MBHs pairs that co-evolve during galaxy mergers. The pairings of MBHs depend on cosmological factors such as hierarchical structure formation, and the large-scale dynamics of MBHs during galaxy mergers, which determines what pairs form bona-fide MBHBs. While the initial mass ratios can be estimated from cosmological simulations, accretion episodes following a galaxy merger are expected to modify this distribution \citep[e.g.,][]{Siwek2020}. Since cosmological simulations have insufficient resolution to track the accretion and evolution of MBHBs at sub-kpc scales, these populations must be modeled through semi-analytic methods. 

MBH pairs may accrete independently from one another at kpc scales, and their mass ratios at binary formation will depend on how much gas has been stripped from or retained in the vicinity of the respective MBHs following a galaxy merger. Due to the resolution limitations discussed above, it is currently unclear how accretion prior to binary formation affects the mass ratio distribution.
At sub-parsec separations however, CBD-driven accretion models can be used to model the mass ratio evolution. The processes determining gas inflow rates to the galactic nucleus are determined on larger scales than the binary separation, such that only the total binary mass is relevant, and a total accretion rate onto the binary is estimated. CBD simulations can then be utilized to divide the accretion rate between the two black holes, evolving the mass ratio of the system. 

\cite{Siwek2020} showed that preferential accretion onto the secondary MBH shifts the mass ratio distribution towards unity, and depletes the number of lower mass ratio systems. This has significant consequences for multi-messenger studies of MBHBs, modifying the population statistics of transients, their merger rates and the gravitational wave signatures emitted during inspiral. In this work we verify these results with updated CBD accretion models \citep{Siwek2023a} that account for the effect of orbital eccentricity on the binary mass ratio evolution, and in turn the effect of the evolving mass ratio on the semi-major axis and eccentricity evolution rates \citep{Siwek2023b}, which in turn determines the abundance of sub-parsec systems per mass ratio bin.

While initial mass ratio distributions can be constrained by cosmological simulations and semi-analytic population studies (e.g. \citealt{Siwek2020}), the evolution of orbital eccentricity of MBHs prior to binary formation is highly uncertain. 
At kpc scales, recent simulations have found that MBHs may be on highly eccentric orbits around the post-merger galactic center ($e_{\rm b} \sim 0.7$; e.g., \citealt{Chen2022}). \cite{Gualandris2022} analyzed massive gas-free galaxy mergers from the early stages of a merger to the hardening phase, and similarly found the orbital eccentricity of the MBHB at formation to be highly eccentric, in the range $0.5 \lesssim e_{\rm b} \lesssim 0.9$. 
The eccentricity of MBHBs also depends on their orbital motion in the core of their host galaxies. In particular, the eccentricity evolution is sensitive to the binary motion relative to the co-rotating or counter-rotating stellar background. Generally, co-rotation results in low orbital eccentricities \citep[e.g.,][]{Dotti2007}, while high eccentricities have been seen in counter-rotating models \citep[e.g.,][]{Sesana2011, MadiganLevin2012, Mirza2017}.
However, small perturbations in the galaxy merger orbit can lead to large variations in the MBHB eccentricity at formation \citep{Rawlings2023}, highlighting that the initial eccentricity distribution of MBHBs at formation is still highly uncertain.

As the two MBHs become gravitationally bound, their orbital evolution is governed by angular momentum exchange with individual stars. While ejecting stars from the galactic center, the semi-major axis decreases and the orbital eccentricity increases \citep{Quinlan1997,Sesana2006}. \cite{Sesana2010} found that residual eccentricities from stellar scattering may be detectable in the LISA and PTA bands, ranging from $10^{-3} \rightarrow 0.2$ and $0.03 \rightarrow 0.3$ respectively (see also \citealt{Porter_Sesana2010} who find even higher eccentricities as binaries enter the LISA band). 
However, the effect of stellar scattering on orbital eccentricity is mild, dependent on the uncertain initial eccentricity distribution set at large scales.

The initial eccentricity and mass ratio distributions of MBHBs at formation are therefore highly uncertain.
In this paper we apply the most recent CBD driven evolution models of binaries from \cite{Siwek2023a,Siwek2023b} to a sample of MBHBs from the \texttt{Illustris} simulation. By analyzing the MBHB population in the observable multi-messenger bands, we show that: \begin{enumerate*}
   \item the CBD erases the initial conditions of the highly uncertain binary eccentricity distribution by quickly and efficiently evolving the eccentricity to an equilibrium value,
   \item unless all binaries form on extremely eccentric orbits ($e_{\rm b, 0} \gtrsim 0.9$), eccentricities in CBD driven systems are up to 2 orders of magntiude higher than in gas-poor systems,  
   \item mass ratio distributions are shifted towards unity in CBD-driven evolution models, and
   \item the GWB amplitude resulting from the MBHB population is boosted by $100\%$ when binaries accrete at just $10\%$ their Eddington limit. 
\end{enumerate*}

\section{Methods}
\label{sec:methods}
%\subsection{\texttt{holodeck}}
%\label{subsec:holodeck}
We generate a sample of MBHBs from the \texttt{Illustris} simulation \citep{Vogelsberger2014b, Vogelsberger2014a, Genel2014} and semi-analytically evolve the population over several evolution (``hardening") regimes using the MBHB population synthesis code \texttt{holodeck} (\citealt{Kelley2017MNRAS}, Kelley et al. in prep.).  Similar to previous work in \cite{Kelley2016, Kelley2017MNRAS, Kelley2017, Siwek2020}, binaries are evolved with several hardening mechanisms \citep{Begelman1980}:
\begin{enumerate}
    \item Dynamical friction (DF) prescription \citep{Begelman1980}
    \item Stellar Scattering (SS) \citep{Quinlan1997, Sesana2006}
    \item CBD torques and accretion \citep{Siwek2023a, Siwek2023b}
    \item GW emission \citep{Peters1964}
\end{enumerate}

The addition of CBD torques and accretion is the main focus of this work. We implement cutting-edge binary evolution models from \cite{Siwek2023a,Siwek2023b} and apply these to MBHBs that have reached the final parsec within a Hubble time. Our model includes evolution rates for binary mass ratio $\dot{q}_{\rm b}(q_{\rm b}, e_{\rm b})$, semi-major axis $\dot{a}_{\rm b}(q_{\rm b}, e_{\rm b})$ and orbital eccentricity $\dot{e}_{\rm b}(q_{\rm b}, e_{\rm b})$.
As shown in \cite{Siwek2023a, Siwek2023b}, the evolution rates depend on both binary mass ratio and eccentricity simultaneously, and thus we interpolate the evolution rates as the binary systems evolve. CBD-driven evolution furthermore depends on the accretion rate of the binary system, which we parameterize by the Eddington fraction $f_{\rm Edd}$. In this work, we fiducially assume a constant accretion rate at 10\% of the Eddington limit for each binary system, $f_{\rm Edd} = 0.1$. We assume that CBDs form around binaries upon reaching parsec separation, as AGN disks at larger than parsec scales may be susceptible to disk fragmentation \citep{Goodman2003MNRAS.339..937G}. The CBD-driven accretion and orbital evolution models are therefore applied in each binary system from parsec scales until merger. To motivate this assumption, we refer to the AGN disk lifetime, which is expected to persist for around $\sim 10^8$ years \citep[e.g.,][]{YuTremaine2002}, and up to $10^9$ years for some models \citep{Marconi2004MNRAS.351..169M}. As these timescales are comparable to the binary hardening timescale at parsec separation \citep{Kelley2017}, we assume that the CBD is long-lived and remains around the binary throughout its evolution.

We investigate two fiducial scenarios. Initially, each MBHB forms on an almost circular orbit with a small initial eccentricity $e_{\rm b,0} = 0.01$, and is then evolved with or without CBD-driven orbital evolution and accretion. We choose initially nearly circular orbits in order to compare our results with previous studies, where circular MBHBs are typically assumed (e.g., \citealt{Katz2020, Bortolas2021}).

\begin{enumerate}
    \item \texttt{nCBD_eb0=1.e-2}: Binaries form with an initial eccentricity $e_{\rm b, 0} = 10^{-2}$ and experience DF, Stellar Scattering and GW evolution. We apply no accretion, i.e. $\dot{M}_{\rm b} = 0$ and as a result, CBD-driven evolution rates are also zero, $\dot{a}_{\rm b,CBD} = 0$ and $\dot{e}_{\rm b,CBD} = 0$.
    \item \texttt{yCBD_eb0=1.e-2}: Binaries form with an initial eccentricity $e_{\rm b, 0} = 10^{-2}$, and subsequently experience DF, SS, CBD and GW evolution. Our fiducial accretion rate is set to 10\% Eddington, and as a result, accretion and CBD-driven evolution rates are non-zero, i.e. we apply $\dot{M}_{\rm b} = 0.1\, \dot{M}_{\rm b,Edd}$, $\dot{a}_{\rm b,CBD}(q_{\rm b}, e_{\rm b})$ and $\dot{e}_{\rm b,CBD}(q_{\rm b}, e_{\rm b})$.
\end{enumerate}

To account for uncertainties in the initial eccentricity of binary systems at formation, we investigate two additional scenarios where binaries form with a random initial eccentricity:

\begin{enumerate}
    \item \texttt{nCBD_eb0=rand}: Binaries form with a random initial eccentricity between $0 \rightarrow 0.999$, and subsequently experience DF, SS and GW evolution. We apply no accretion, i.e. $\dot{M}_{\rm b} = 0$ and as a result, CBD-driven evolution rates are also zero, $\dot{a}_{\rm b,CBD} = 0$ and $\dot{e}_{\rm b,CBD} = 0$.
    \item \texttt{yCBD_eb0=rand}: Binaries form with a random initial eccentricity between $0 \rightarrow 0.999$, and subsequently experience DF, SS, CBD and GW evolution. Our fiducial accretion rate is set to 10\% Eddington, and as a result, accretion and CBD-driven evolution rates are non-zero, i.e. we apply $\dot{M}_{\rm b} = 0.1\, \dot{M}_{\rm b,Edd}$, $\dot{a}_{\rm b,CBD}(q_{\rm b}, e_{\rm b})$ and $\dot{e}_{\rm b,CBD}(q_{\rm b}, e_{\rm b})$.
\end{enumerate}

Additionally, we investigate the effect of accretion rates on the population by probing regimes where binaries accrete at Eddington fractions $f_{\rm Edd} = [0.01, \ 0.10, \ 1.00]$. For each population, we report eccentricity and mass ratio distributions in the LSST and PTA bands, GWB amplitudes, and detection rates in the LISA band.
In the following we analyze the results and report the effect of CBD-driven evolution on properties of the MBHB population. 

\section{Results}
\subsection{Eccentricity Evolution in MBHB Populations}
\begin{figure}
    \centering
\includegraphics[width=1.0\columnwidth]{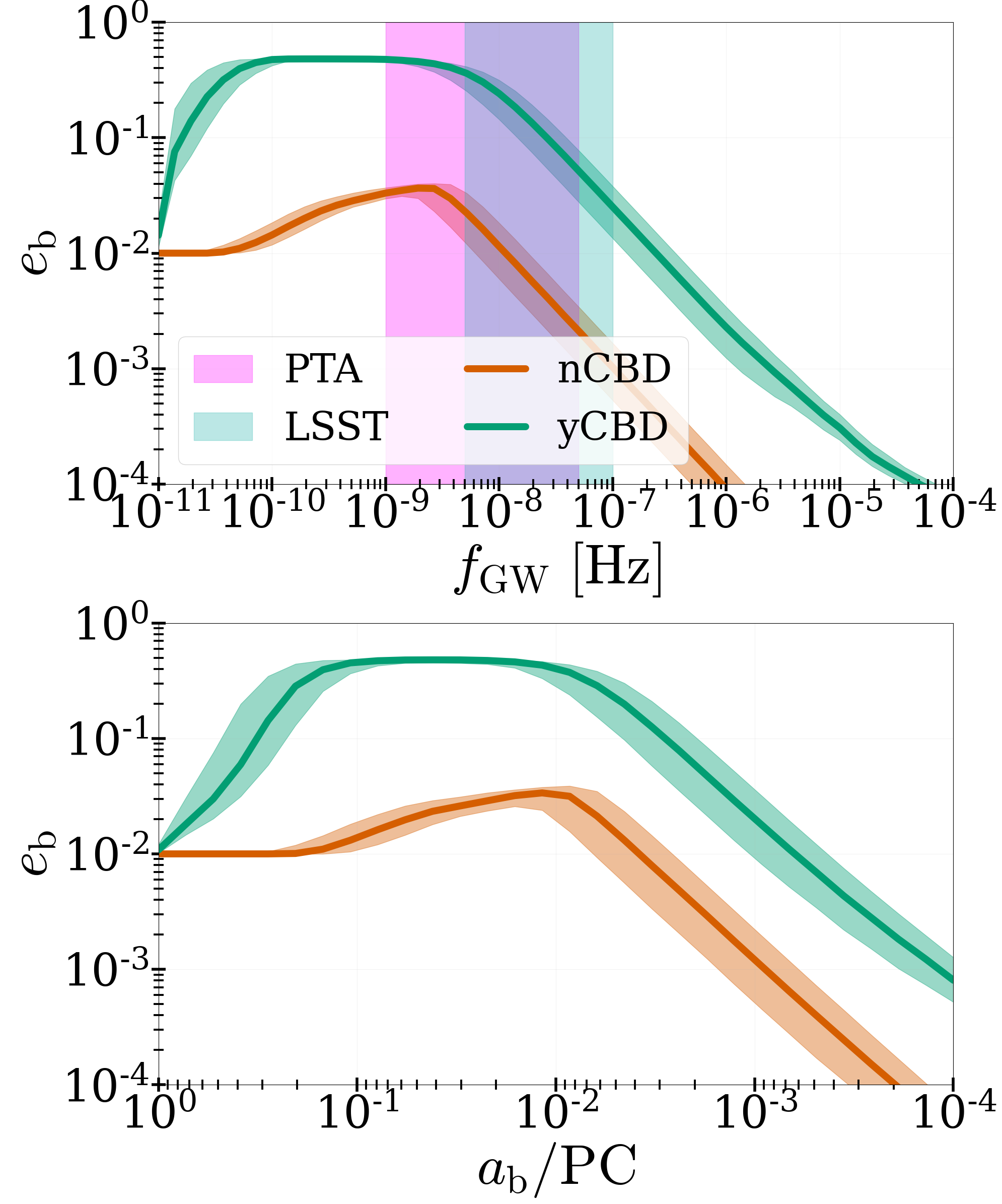}
    \caption{Median, 32-nd and 68-th percentile of binary eccentricity as a function of frequency (top panel) and semi-major axis (bottom panel), showing fiducial runs \texttt{yCBD_eb0=1.e-2} and \texttt{nCBD_eb0=1.e-2} (green and orange lines, respectively). CBD torques boost the median eccentricities of close binaries by 1-2 orders of magnitude. Significant eccentricities are expected in sub-parsec MBHBs, as they enter the PTA band. Large orbital eccentricities are still present in the LSST band, where GW-driven circularization begins to take over.}
\label{fig:freq_sepa_eb}
\end{figure}

In Figure \ref{fig:freq_sepa_eb} we show the median binary eccentricity as a function of binary separation (top panel) and GW frequency (bottom panel) for all binaries in our sample. The shaded regions around each line correspond to the 32-nd and 68-th percentiles. We compare our two fiducial models: Binaries initialized at a low eccentricity $e_{\rm b,0} = 0.01$, and evolved either without CBD torques and accretion (orange line), or with both CBD torques and preferential accretion models \citep[][green line]{Siwek2023a,Siwek2023b}. While both models start out at the same initial eccentricity $e_{\rm b,0} = 0.01$ at $a_{\rm b} = 1\,\rm{pc}$, CBD torques lead to a sharp increase in binary eccentricity, up to a median peak eccentricity capped at $\sim 0.5$. This is in line with the CBD induced equilibrium binary eccentricity recently reported in \cite{Zrake2021, DorazioDuffell2021, Siwek2023b}. As GW emission becomes the dominant evolution process, MBHBs begin to circularize at $a_{\rm b} \sim 10^{-2} \, \rm{PC}$, or $f_{\rm GW} \sim 10^{-8}\, \rm{Hz}$. However, since the GW circularization rate is the same for both populations, a 1-2 order of magnitude gap remains that distinguishes the CBD driven model from the model which excluded CBD torques. 

\begin{figure}
    \centering
\includegraphics[width=1.0\columnwidth]{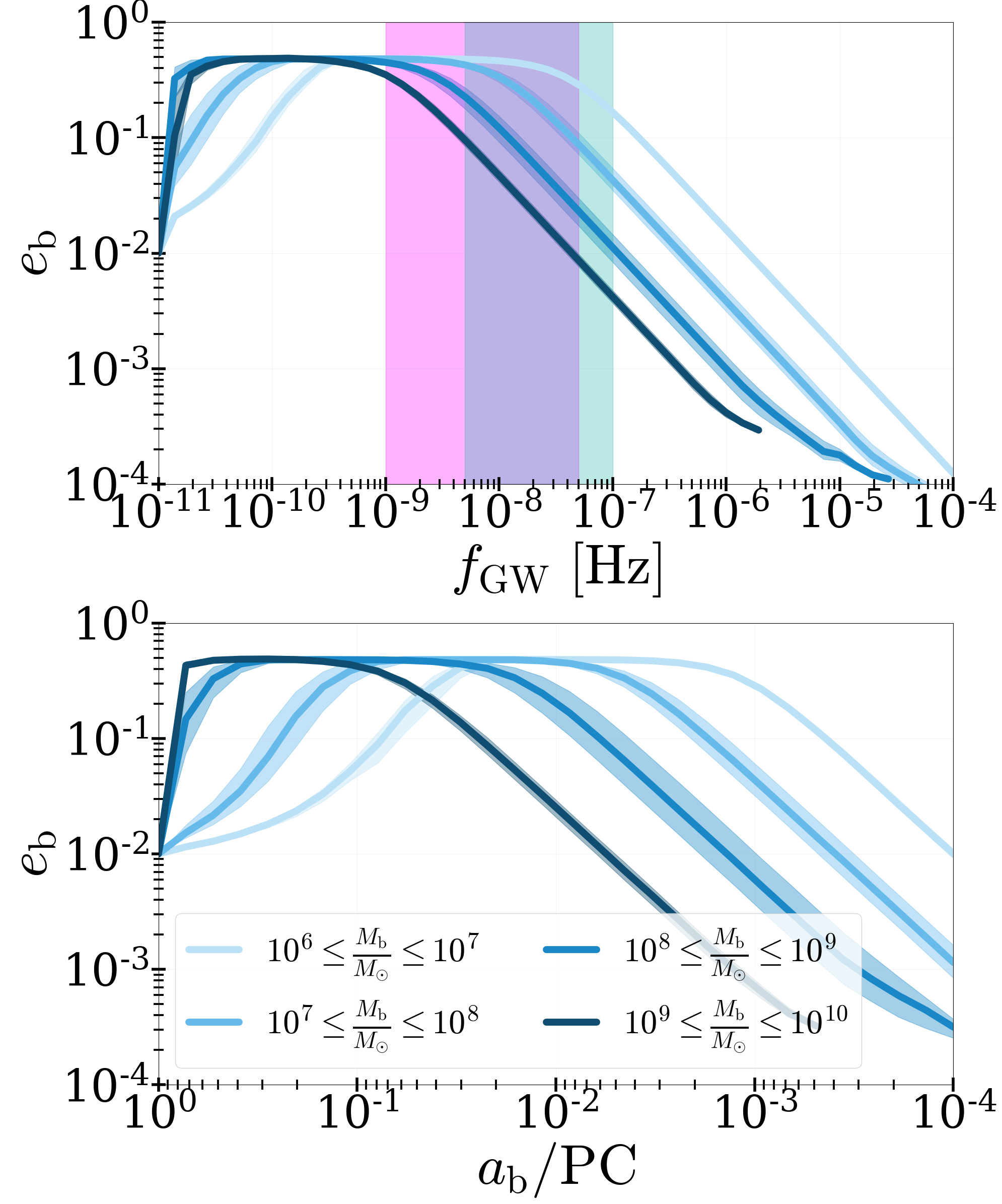}
    \caption{MBHB orbital eccentricity evolution in our fiducial \texttt{yCBD_eb0=1.e-2} simulation. Median binary eccentricity along with 32-nd and 68-th percentiles are shown as a function of GW frequency (top panel), again showing the PTA and LSST regimes in pink and teal respectively. The bottom panel shows the orbital eccentricity as a function of semi-major axis. Here, the MBHB population is separated into mass bins to analyze the mass dependence of the eccentricity evolution at a given frequency/separation. }
\label{fig:freq_sepa_eb_mass_binned}
\end{figure}

In Figure \ref{fig:freq_sepa_eb_mass_binned} we show the median, 32-nd and 68-th percentile of CBD driven binary eccentricity as a function of GW frequency (top panel) and binary separation (bottom panel) for a range of mass bins: $10^6 \, M_{\odot} \lesssim M_{\rm b} \lesssim 10^7 \, M_{\odot}$, $10^7 \, M_{\odot} \lesssim M_{\rm b} \lesssim 10^8 \, M_{\odot}$, $10^8 \, M_{\odot} \lesssim M_{\rm b} \lesssim 10^9 \, M_{\odot}$ and $10^9 \, M_{\odot} \lesssim M_{\rm b} \lesssim 10^{10} \, M_{\odot}$, indicated by colors ranging from light to dark blue. 
Similar to the green line in Figure \ref{fig:freq_sepa_eb}, eccentricity increases as a result of CBD-driven orbital evolution before circularization due to GWs takes over. The transition frequency/separation depends on total binary mass: at higher binary masses, GW dominated evolution takes over earlier than it does in lower mass binaries. We attribute this to the scale-free nature of the hydrodynamic simulations that our CBD-driven evolution models are based on. While $\dot{e}_{\rm b}$ and $\dot{a}_{\rm b}$ are independent of binary mass, the GW-driven orbital evolution rates do depend on binary mass. As a result, the transition from CBD-driven to GW-driven evolution occurs earlier for higher mass binaries and later for lower mass binaries. This leads binaries with lower masses to retain higher eccentricities than higher mass binaries, when compared at the same separation. 

\begin{figure}
    \centering
\includegraphics[width=1.0\columnwidth]{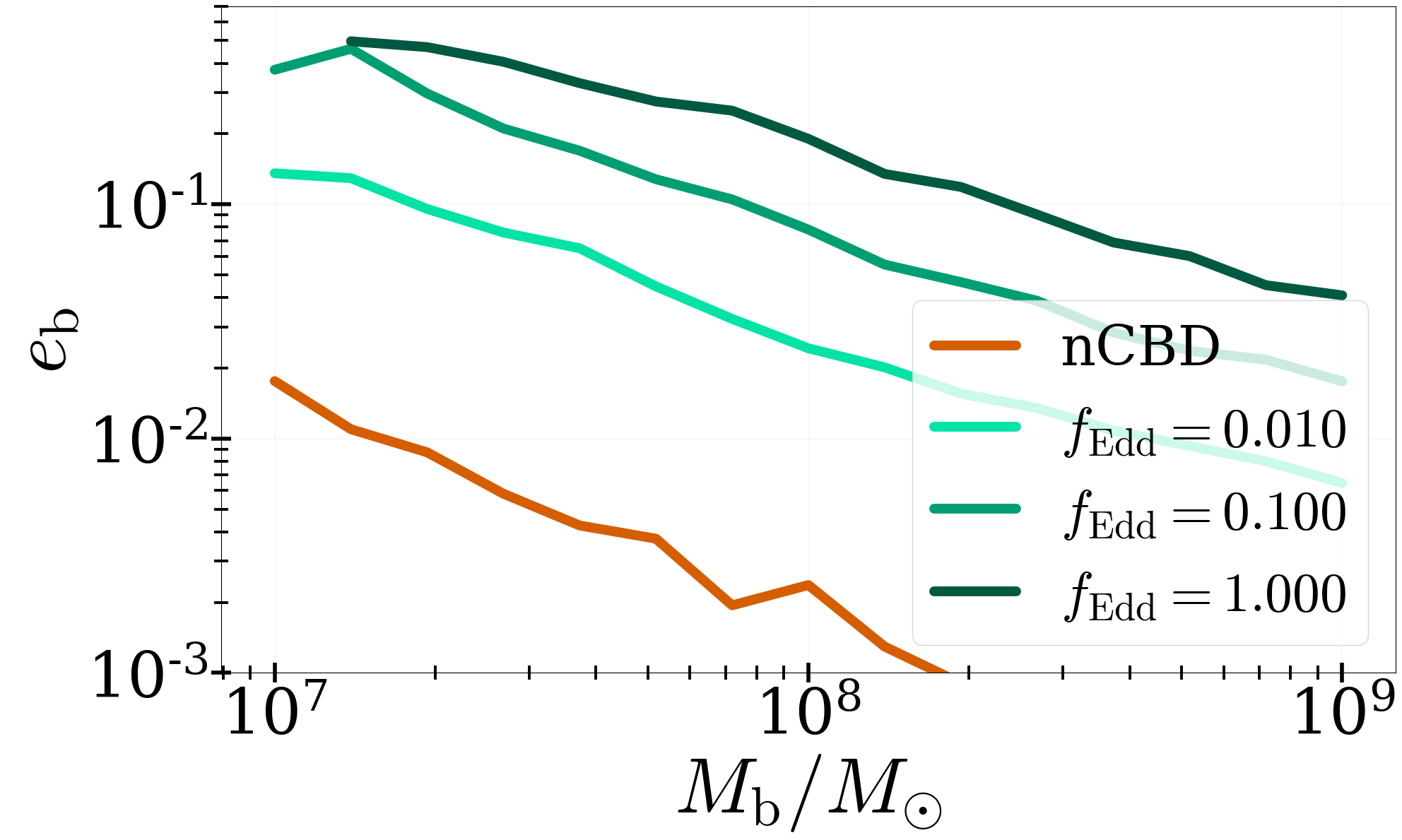}
    \caption{Orbital eccentricity versus combined mass in binary systems within the LSST and PTA frequency bands. The orange line represents a population of MBHBs not evolved with CBD physics, while light, medium and dark greens refer to MBHB populations evolved with CBD physics and increasing accretion rates. In all cases, binary mass is associated inversely with orbital eccentricity. }
\label{fig:eb_vs_mb_PTA_LSST_mb_eb}
\end{figure}

In Figure \ref{fig:eb_vs_mb_PTA_LSST_mb_eb} we show MBHB populations in the LSST and PTA bands, plotting the median binary eccentricity as a function of binary mass. We contrast populations evolved without CBD physics (orange line), and with CBD physics applied at increasing accretion rates. The light green line indicates that all binaries accrete at 1\% of their Eddington limit, medium green corresponds to 10\% Eddington and dark green to 100\%. In all cases, the median orbital eccentricity is inversely related to binary mass. The slopes in all cases are approximately similar, while accretion rate increases the amplitude of the $e_{\rm b}$ versus $M_{\rm b}$ relation.

\subsection{Dependence on initial conditions}
\begin{figure}
    \centering
\includegraphics[width=1.0\columnwidth]{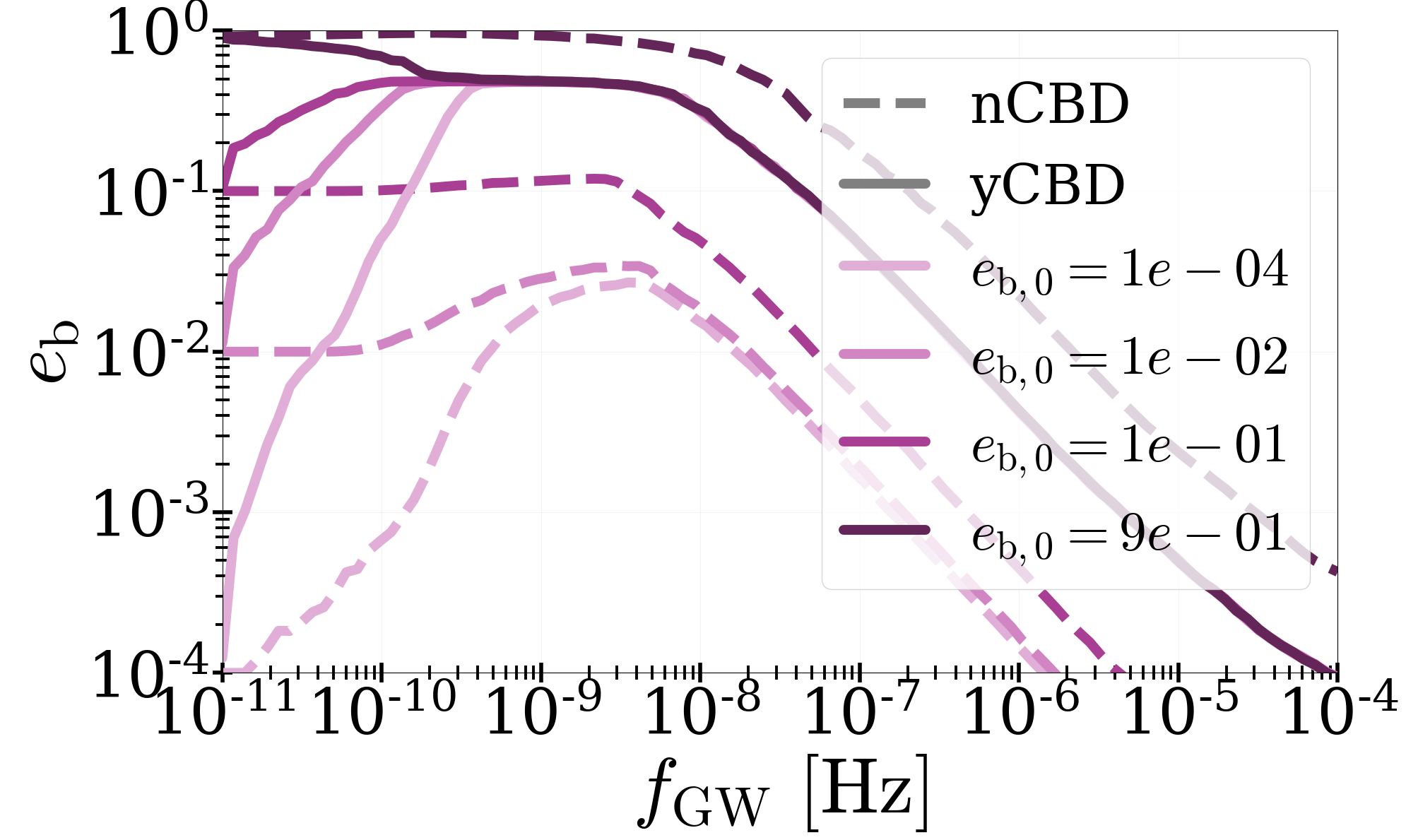}
    \caption{Median eccentricities of MBHBs as a function of GW frequency, comparing a range of initial eccentricities. The solid and dashed lines show populations evolved with and without CBD physics respectively, while colours indicate the orbital eccentricity at which the binary systems were initialized. The convergence of the solid lines indicates that CBD physics erases the initial conditions and dominates the eccentricity evolution of sub-parsec MBHBs, despite initial eccentricities varying by several orders of magnitude.}
    \label{fig:init_eb0_gw_freq_eb}
\end{figure}

The initial eccentricity of MBHBs is intrinsically uncertain, since even small perturbations in the galaxy merger orbit can lead to large variations in the final MBHB eccentricity \citep[e.g.,][]{Rawlings2023}. 
To check the dependence of a potentially observable eccentricity distribution on the initial binary eccentricity, we run several realizations of our binary evolution with a range of initial eccentricities: $e_{\rm b, 0} = [0.0001, 0.01, 0.1, 0.9]$, and compare the eccentricity evolution as a function of binary separation in each case. The results are shown in Figure \ref{fig:init_eb0_gw_freq_eb}. In the case of binaries evolved without CBD torques (dashed lines), the initial eccentricity of the binary strongly influences the population going forward. However, when we apply CBD torques (solid lines), the initial conditions are erased: all binaries end up at the same eccentricity-separation track. The CBD evolved population maintains a higher orbital eccentricity at sub-parsec separations than any other realization without CBD effects, unless all binaries are initialized with a very high initial eccentricity ($e_{\rm b,0} \sim 0.9$). Our results therefore indicate that, unless binaries form at extremely high initial eccentricities, CBD evolved populations will achieve the highest median orbital eccentricities at sub-parsec separations, and the value of this eccentricity is always the same, irrespective of the initial conditions. Unless other eccentricity pumping mechanisms are invoked (e.g. three body interactions, Kozai-Lidov), CBD torques are the only mechanism leading to the MBHB eccentricity distributions we report here.

\subsection{CBD Signatures in MBHB Populations}
\label{sec:cbd_signatures}
PTAs have recently found strong evidence for a Gravitational Wave Background (GWB) emitted by MBHBs \citep[e.g.,][]{NANOGravDetection2023, NANOGravMBHBs2023} during their long-term inspiral. Such a measurement probes the net MBHB population, and can constrain their number densities and masses. Additionally, PTAs are expected to detect individual MBHBs as ``Continuous Wave" (CW) sources \citep{Kelley2018MNRAS.477..964K}, which would allow for detailed study of sub-parsec MBHB dynamics without the degeneracies of modeling large populations. A recent analysis of the NANOGrav 15-year data set yielded only upper limits, but some tantalizing hints of excess power at $\sim 4 \, \rm{nHz}$ and $\sim 170 \, \rm{nHz}$ \citep{NANOGrav2023_CW}.

A much larger sample of MBHB candidates is likely to be detected \citep{Kelley2019} by the Legacy Survey of Space and Time (LSST; \citealt{LSST2019}) at the Rubin Observatory \citep{LSST2019}. In addition to PTAs probing the MBHB population via GW detection, LSST will reveal the EM counterparts of such systems in the optical band. This will enable a multi-messenger view of the evolution and fate of MBHBs across cosmic time. LSST is expected to detect up to $10^8$ MBHBs \citep{Xin2021}, providing a detailed population overview up to redshift $z \sim 6$. 

For the purposes of this analysis, we consider MBHBs in the PTA range emitting at frequencies between $10^{-9}\, \rm{Hz}$ to $ 10^{-8}\,\rm Hz$. We sample the MBHB population generated as stated in Section \ref{sec:methods} in this frequency range to obtain characteristics of this population, including their masses, mass ratios and eccentricities.
LSST will nominally observe for $10$ years \citep{LSST2019}, and will require at least 1 binary period to identify variability in a binary AGN candidate. Conservatively, we select only binaries in our sample that can complete at least 2 orbital cycles over LSST mission time, and fall below the higher frequency cadence of $\sim$ 1 week. With those parameters we define the LSST frequency band $6 \times 10^{-9} \rm{Hz} \ \rightarrow 10^{-6} \  \rm Hz$. 
The orbital frequencies corresponding to the PTA band overlap significantly with the range of orbital frequencies detectable with LSST (see also the shaded regions in Figure \ref{fig:freq_sepa_eb}). As a result, the population statistics are similar, and we show the combined PTA and LSST detectable populations in our results. 

\begin{figure*}
    \centering
\includegraphics[width=1.0\textwidth]{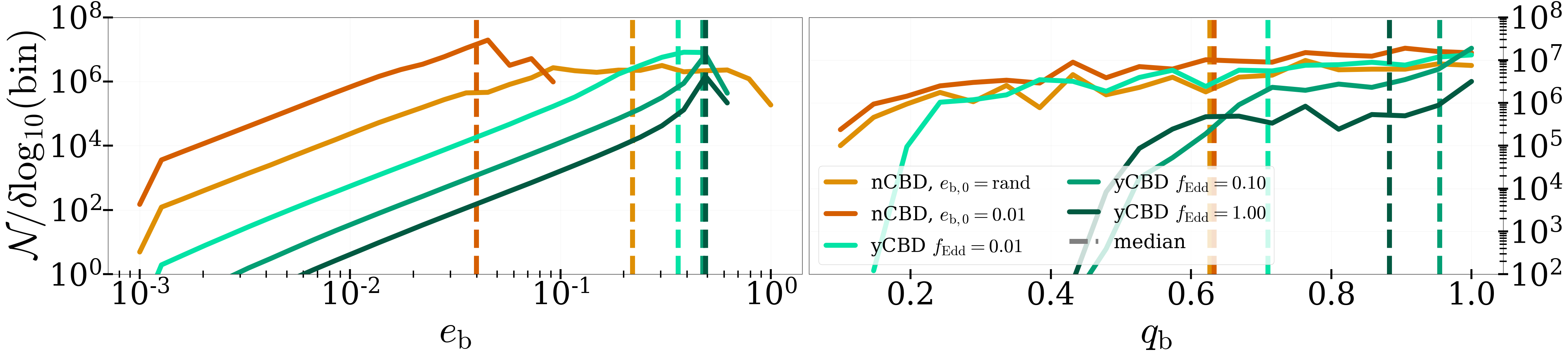}
    \caption{Number of MBHBs in an observer's lightcone as a function of binary eccentricity (left) and mass ratio (right), including binaries in the PTA and LSST frequency bands. The dark and light orange lines show models evolved without CBD torques, with either $e_{\rm b,0} = 0.01$ or random initial eccentricities, respectively. The green lines show populations evolved with CBD models from \protect\cite{Siwek2023a,Siwek2023b}. The shade of green refers to the Eddington fraction at which binaries accrete, with the lightest green representing 1\% Eddington and the dark green 100\% Eddington. In all cases, CBD evolved populations have higher median eccentricities and mass ratios than populations evolved without CBD torques.}
\label{fig:eb_qb_dist}
\end{figure*}

In the following we present the characteristics of the MBHB populations in our sample that fall into the PTAs and LSST frequency bands. We show all the binaries in our sample that fall into the frequency range of PTAs and LSST. We do not take any luminosity cuts to account for LSST sensitivity. This is because luminosity cuts turn out to be irrelevant for binaries accreting at 10\% Eddington, as most of these systems are luminous enough to be detected, and because we aim to study the underlying MBHB population, without considering detection bias.
In Figure \ref{fig:eb_qb_dist} we show the expected number of MBHBs at any given time (per logarithmic bin width) as a function of binary eccentricity ($e_{\rm b}$; left) and mass ratio ($q_{\rm b}$; right). We contrast distributions from populations evolved without CBD models (orange lines) and with CBD accretion and orbital evolution models from \cite{Siwek2023a, Siwek2023b} (green lines). Most populations start with the fiducial initial binary eccentricity $e_{\rm b,0} = 0.01$, aside from the light orange line which is evolved without CBD physics but initialized with a random eccentricity distribution. 
The shade of the green line refers to the accretion rate parameterized by the Eddington accretion rate fraction, which increases from 1\% (light green) and 10\% (fiducial; medium green), up to 100\% (dark green). 
MBHBs evolved with CBD models enter the PTA/LSST frequency range with significantly higher eccentricities. The median eccentricity in the case without CBDs, initialized at $e_{\rm b,0} = 0.01$, is at $e_{\rm b,med} \sim 4\times 10^{-2}$, just above the initial eccentricity $e_{\rm b,0} = 0.01$ (dark orange line). In the case of the \texttt{nCBD} population evolved with random initial eccentricity (light orange), the median eccentricity in the PTA/LSST band is higher, at $e_{\rm b,med} \sim 2\times 10^{-1}$, but still lower than the median eccentricities reached by any of the \texttt{yCBD} cases. The populations evolved with CBD torques measure median eccentricities up to $e_{\rm b,m} \sim 0.5$, and sharply decrease in number density below the peak at $e_{\rm b,m} \sim 0.5$. 

In the right panel, we show the mass ratio distributions for all our models using the same colours as before. 
We find that the mass ratio distribution in the populations evolved without CBD accretion or with very low accretion rate (1\%, light green line) is flat, with a median $q_{\rm b,m} \sim 0.6$. In the \texttt{yCBD} cases (medium and dark green lines), the mass ratio distribution is sharply leaked towards unity, and median mass ratios are  $q_{\rm b,m} \gtrsim 0.9$. 

\begin{figure*}
    \centering
\includegraphics[width=1.0\textwidth]{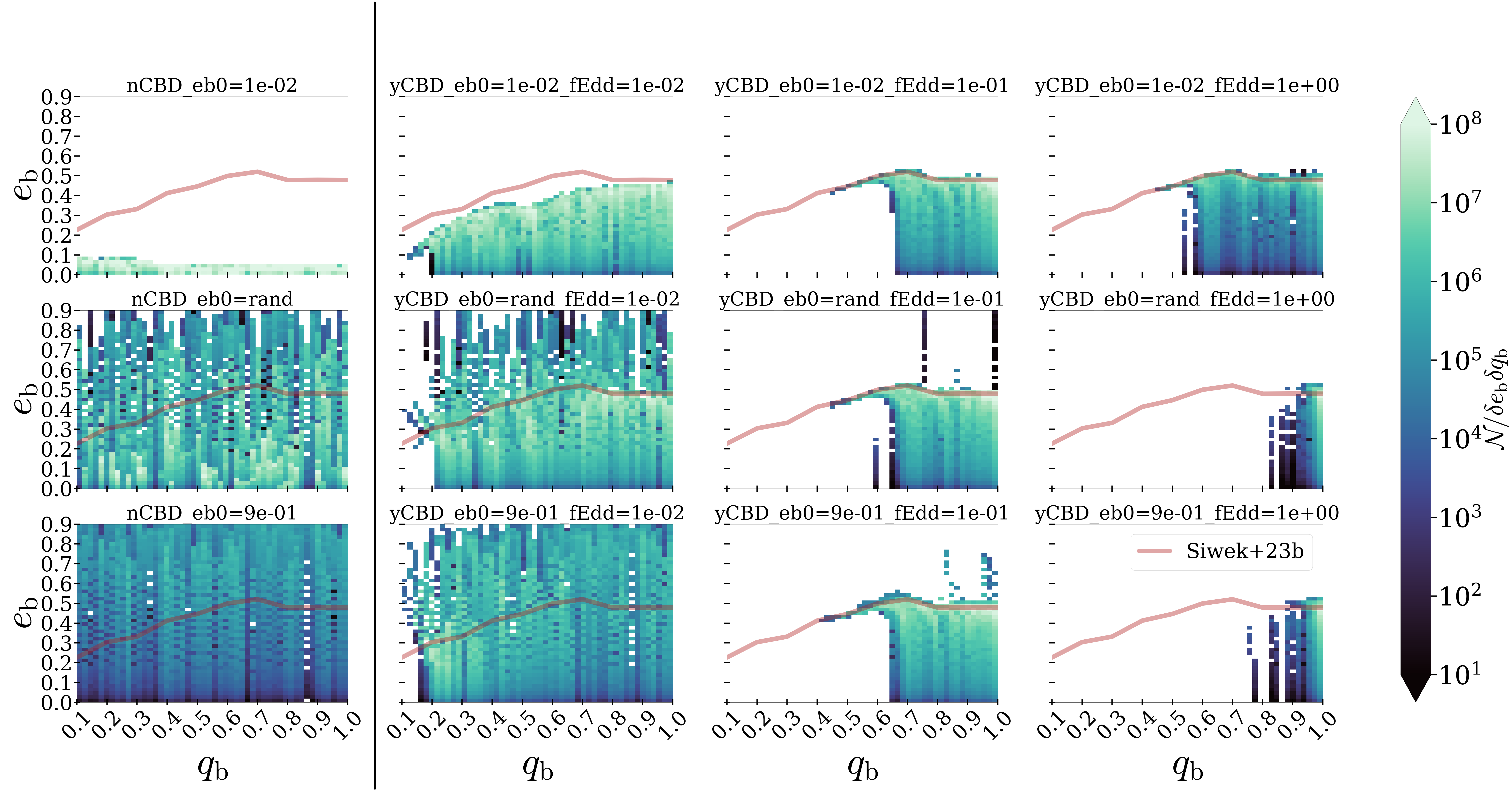}
    \caption{Eccentricities of MBHBs as a function of binary mass ratio, corresponding to binaries in the PTA and LSST frequency ranges. The red line shows binary equilibrium eccentricities as a function of mass ratio, as found in \protect\cite{Siwek2023b}. The first column shows MBHBs evolved without CBDs. The following columns going from left to right show MBHBs evolved with CBDs, with increasing accretion rates from 1\% Eddington (first column after vertical line), up to 100\% Eddington in the right-most column. The rows indicate the initial eccentricity distribution of MBHBs at formation, with the top row showing our fiducial value $e_{\rm b,0} = 0.01$, the middle row uses a randomly drawn initial eccentricity distribution, and the bottom row sets $e_{\rm b,0} = 0.9$ uniformly.}
    \label{fig:hmap_LSST_PTA}
\end{figure*}

In Figure \ref{fig:hmap_LSST_PTA} we show the PTA/LSST MBHB population across the mass ratio versus eccentricity parameter space. We contrast simulations with various initial conditions, evolution models and accretion parameters: In the first column, separated from the rest of the panels, we show the populations evolved without CBD torques. We test 3 different initial eccentricity distributions: $e_{\rm b,0} = 0.01$ (top panel), randomly drawn values of $e_{\rm b,0}$ in the range $[0,1.0)$ (middle panel) and $e_{\rm b,0} = 0.9$ (bottom). We find that significant memory of the initial eccentricity distribution is retained in the populations when no CBD torques are included.

The remainder of the panels on the right hand side of the black bar show MBHB populations evolved with CBD physics. The columns increase from left to right in their accretion rates. The first column shows binary systems with $e_{\rm b,0} = 0.01$ (top), $e_{\rm b,0} = \rm{rand}$ (middle) and $e_{\rm b,0} = 0.9$ (bottom), all evolved with CBD physics but at low accretion rates $f_{\rm Edd} = 0.01$. While the initial eccentricity distribution affects the population statistics, the systems have evolved towards the equilibrium eccentricity $e_{\rm b,eq}$ versus $q_{\rm b}$ relation, overplotted in the red line. The second column shows populations again with $e_{\rm b,0} = 0.01$ (top), $e_{\rm b,0} = \rm{rand}$ (middle) and $e_{\rm b,0} = 0.9$ (bottom), but evolved at the fiducial accretion rate $f_{\rm Edd} = 0.10$. In this case, the population statistics are nearly identical, and closely trace the $e_{\rm b,eq}$ versus $q_{\rm b}$ relation. Any initial conditions appear to have been erased by the CBD-driven evolution.
In the bottom row, the accretion rate is set to $f_{\rm Edd} = 1$. In this case, most binaries have evolved towards $q_{\rm b} = 1$, except in the case where $e_{\rm b,0} = 0.01$, where a low mass ratio tail remains. We suggest that this is due to an increased number of low mass ratio binaries reaching the PTA/LSST bands as a result of stronger CBD-driven inspiral in near-circular systems (see Figure 2 in \citealt{Siwek2023b}). In the middle and bottom panels, where $e_{\rm b,0} = \rm{rand}$ (middle) and $e_{\rm b,0} = 0.9$ (bottom), most binaries have equal mass ratios and eccentricities $e_{\rm b} \lesssim 0.5$.

\subsection{The effect of CBD accretion on the GWB spectrum}
\begin{figure*}
    \centering
\includegraphics[width=1.0\textwidth]{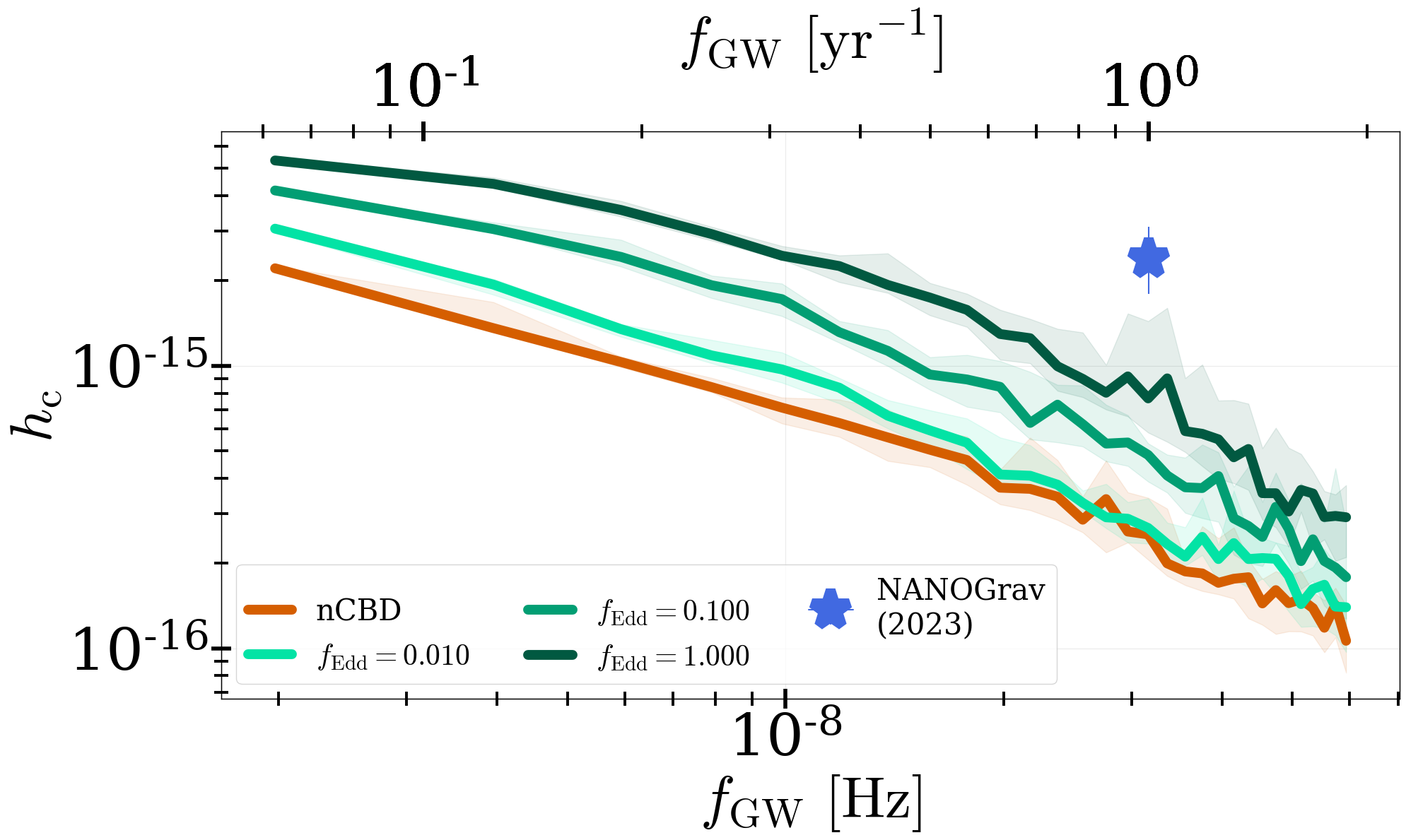}
    \caption{GWB spectrum calculated for models evolved with CBDs, but with different accretion rates. We compare spectra resulting from MBHB populations evolved without CBD physics (orange line) to those with CBD physics at Eddington accretion rates $f_{\rm Edd} = 0.01$ (light green line), $f_{\rm Edd} = 0.10$ (medium green line) and $f_{\rm Edd} = 1.00$ (dark green line).}
\label{fig:gwb_fedd}
\end{figure*}

In Figure \ref{fig:gwb_fedd} we show the GWB spectrum emitted by 4 MBHB populations evolved with distrinct CBD accretion and orbital evolution: the orange line refers to our fiducial \texttt{nCBD\_eb0=1.e-2} model, while the green lines refer to \texttt{yCBD\_eb0=1.e-2}, each evolved with a different accretion rate. The shade of green saturates with increasing Eddington fraction, starting at $f_{\rm Edd} = 0.01$ (light green), $f_{\rm Edd} = 0.10$ (medium green; our fiducial model) and $f_{\rm Edd} = 1.00$ (dark green). 
Each binary in our sample begins accreting once it reaches sub-parsec separation, and subsequently accretes at a constant Eddington fraction. Higher accretion rates evolve MBHBs to overall higher masses, but more importantly also increase the mass ratios. The chirp mass $\mathcal{M}$ of a binary depends sensitively on its mass ratio, 
\begin{equation}
    \mathcal{M} = \Big[ \frac{q_{\rm b}}{(1+q_{\rm b})^2} \Big]^{3/5} \, M_{\rm b}.
\end{equation}
The GW spectrum depends on the chirp mass, and its amplitude $h_{\rm c}$ scales as follows with chirp mass and frequency \citep{Enoki2007},
\begin{equation}
    h_{\rm c}^2 \propto \mathcal{M}^{5/3}\, f^{-4/3}.
\end{equation}
A modest increase in the mass ratio distribution of MBHBs can thus result in a significant increase in the overall GWB amplitude.
We find that even models with just 1\% Eddington accretion throughout the sub-parsec evolution of the binary show a small increase in the GWB amplitude at $f_{\rm GW} = 1\, \rm{yr}^{-1}$. At 10\% Eddington, the GWB amplitude has increased by a factor $2-3$, while at 100\% Eddington the amplitude is over $5$ times higher when compared with the \texttt{nCBD\_eb0=1.e-2} model. 
Our result suggests that CBD accretion can boost the GWB amplitude measurably, and may account for the higher-than-expected amplitude recently inferred by PTAs \citep{NANOGravDetection2023, EPTADetection2023, PPTADetection2023, CPTADetection2023}. We note that the difference in amplitude between our models and the NANOGrav measurement is likely due to an incomplete sample of MBHBs in our simulations. We infer binaries from galaxy mergers in the \texttt{Illustris} simulations, but the volume limit may lead to a dearth in the highest mass MBHs, simply because these systems are already scarce. In future studies, we plan to repeat our population studies with the \texttt{IllustrisTNG} \citep{Marinacci_TNG2018MNRAS.480.5113M, Naiman_TNG2018MNRAS.477.1206N, Nelson_TNG2018MNRAS.475..624N, Pillepich_TNG2018MNRAS.475..648P, Springel_TNG2018MNRAS.475..676S} and \texttt{MillenniumTNG} \citep[e.g.,][]{Delgado_MTNG2023MNRAS.523.5899D} simulations, which cover larger volumes.

We also note that the highest accretion rate models exhibit some indication of a turnover at the lowest frequencies shown here. Current evidence for the GWB suggests that the spectral slope may be flatter than $\propto f^{-2/3}$ as is expected for binaries hardening through GW emission only. The slope carries information about environmental effects and their impact on MBHB dynamics. A flattening of the slope indicates that binaries move through frequency bins faster than through GW emission only, indicating that environmental effects such as gas \citep[e.g.,][]{Kocsis2011} dominate their evolution \citep[see e.g.,][for a review on GWB physics]{Sesana2013CQGra..30v4014S}. Our models provide one possible explanation for this, as higher accretion rates seem to flatten the spectrum out to higher frequencies close to $1\, \rm{yr}^{-1}$.

\subsubsection{CBD Softening: A positive effect on the GWB}
\label{sec:softening}
\begin{figure}
    \centering
\includegraphics[width=1.0\columnwidth]{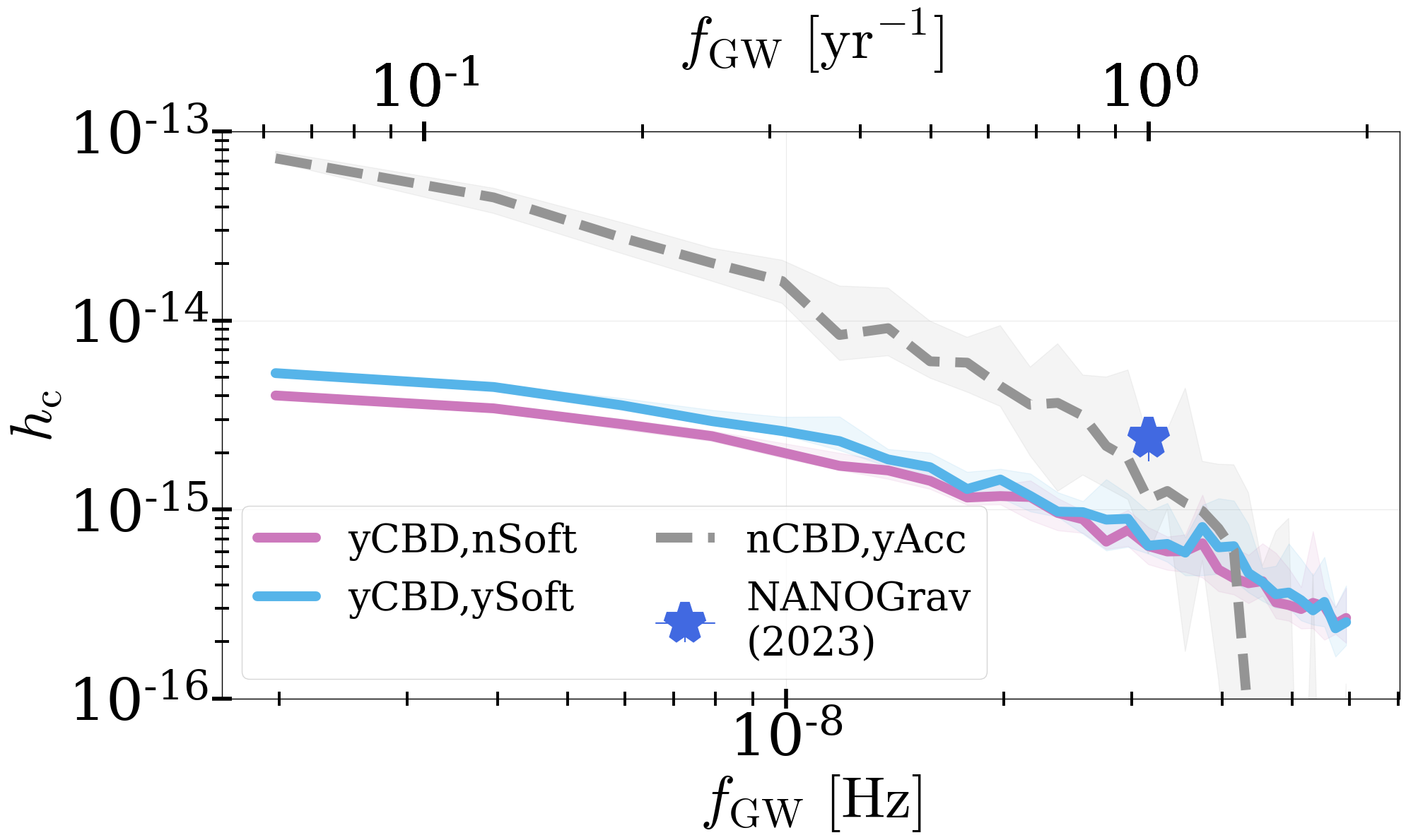}
    \caption{GWB spectrum from MBHB populations evolved with CBD physics, allowing both softening and hardening (blue line) and hardening only (pink line). We also show a model which includes CBD accretion at the same rate as the other two models (100\% Eddington), but does not apply any orbital evolution rates (gray dashed line). This leads to a dramatic increase in the GWB amplitude, as binaries get to grow through accretion without experiencing the hardening effects of the CBD.}
    \label{fig:gwb_softening}
\end{figure}

Recent simulations indicate that the net torque acting between a binary and CBD
can cause orbital expansion \citep[``softening";][]{Miranda2016, Moody2019, Munoz2019, Duffell2020, Munoz2020, DorazioDuffell2021, Siwek2023b}. \cite{Siwek2023b} showed that whether a binary softens or hardens in the presence of a CBD depends on both binary mass ratio and eccentricity. This effect is self-consistently included in our population modeling, by interpolating the hardening/softening rates depending on the orbital parameters of each binary at a given timestep. In Figure \ref{fig:gwb_softening} we show the effect of CBD-driven semi-major axis evolution on the GWB spectrum. We plot the GWB spectrum in a population that includes CBD physics, but neglects any softening rates (i.e., $\dot{M}_{\rm b} > 0$ and $\dot{a}_{\rm b} \leq 0$; pink line) alongside a full CBD model including softening (light blue line). We find that CBD softening results in a slight increase in the GWB amplitude. 
This effect can be attributed to prolonged time periods of accretion. \textbf{Orbital softening allows binaries to accrete over longer time periods ahead of reaching the PTA frequency range}. This allows them to grow to larger masses compared with binaries that do not experience CBD softening. Any softening-induced increase in time spent in the PTA band is not significant in increasing the amplitude, instead, it is the increase in total mass and chirp mass that boosts the GWB amplitude. However, this is a minor effect, and only noticeable if the accretion rates are sufficiently high, here at 100\% Eddington. Other than small increases in total mass, we find no noticeable differences in MBHB population statistics due to CBD softening.

We also investigate a model where binaries accrete gas without any hardening or softening torques from the CBD (gray dashed line in Figure \ref{fig:gwb_softening}). This leads to a dramatic increase in the GWB amplitude, as no hardening effects are applied, and the binaries grow in mass over longer timescales. This limit test serves to account for potentially large uncertainties in CBD evolution models. Current CBD simulations are typically in 2-D, with pure hydrodynamics and an $\alpha$-disk model. However, the semi-major axis hardening rate is quite sensitive to the gas distribution in the CBD cavity, which itself is governed by hydrodynamics in the disk \citep[e.g.,][]{Tiede2020}. As current models neglect significant physical processes in the disk (such as radiative transport, magnetic fields, jets and winds and self-consistent viscous transport through magnetorotational instabilities), the hardening rates from the current 2-D hydrodynamic models are highly uncertain, and could be much smaller than expected. In such a case, the effect on the GWB would again be positive: while binaries take longer to merge, they also have longer timescales to accrete gas from their CBDs and thus boost the GWB amplitude. 

\subsection{MBHBs in LISA}
The Laser Interferometer Space Antenna (LISA; \citealt{AmaroSeoane2017}) will detect the mergers of MBHBs with total masses ranging $10^3 \, M_{\odot} \lesssim M_{\rm b} \lesssim 10^9\, M_{\odot}$ \citep{AmaroSeoane2017, KatzLarson2019}. However, population synthesis models that analyze the detectability of MBHBs typically assume circular orbits \citep[e.g.,][]{Katz2020}. Here we analyze the effect of CBD driven evolution on the detectability of MBHBs in the LISA frequency range. 

\begin{figure*}
    \centering
\includegraphics[width=1.0\textwidth]{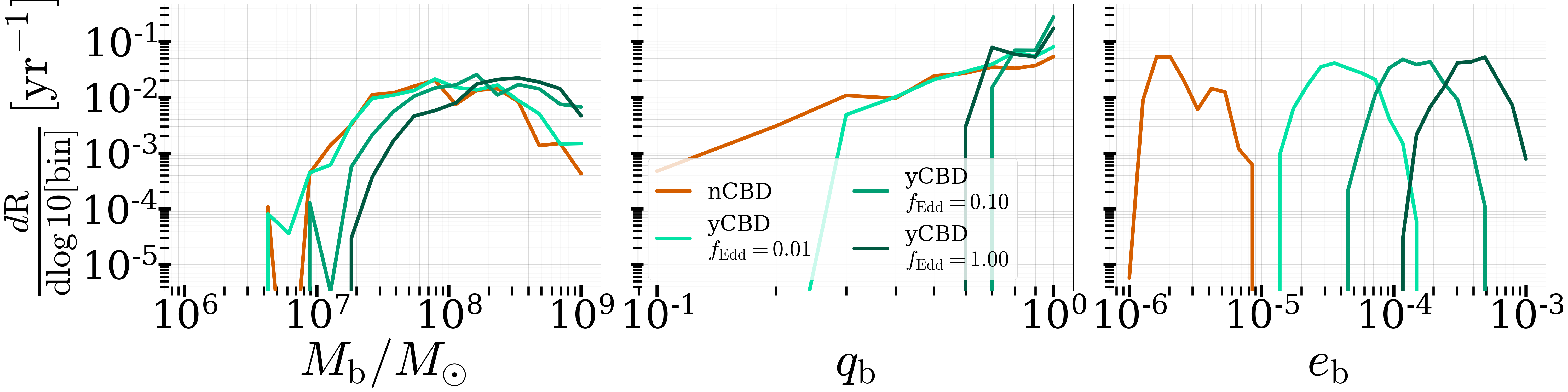}
    \caption{Population statistics of detectable binaries in the LISA band, with rates in each logarithmic bin shown per year. Since our binary sample has a lower mass limit of $\sim 10^6\,M_{\odot}$ at formation, the rates of detections in the LISA band are low.}
\label{fig:lisa_population_fedd_study}
\end{figure*}

Figure \ref{fig:lisa_population_fedd_study} shows the characteristics of the MBHBs in our simulations that could be detected with LISA. To determine whether a source is detectable, we compare its GW strain to the LISA sensitivity at that frequency. We contrast populations evolved without any CBD physics (orange lines) to those evolved with CBD physics to varying magnitude: the pale, medium and dark green lines indicate Eddington accretion rates at 1\%, 10\% and 100\%, respectively. The leftmost panel shows the distribution of total binary masses, with the majority of binaries detectable near $M_{\rm b} \sim 10^8 \, M_{\odot}$. We note that the accretion of gas onto the binary produces only a moderate shift in the total mass distribution, even when binaries accrete at 100\% Eddington throughout their final parsec evolution. 
In the middle panel of Figure \ref{fig:lisa_population_fedd_study} we show the mass ratio distribution in the contrasting populations. Since CBD accretion favours the secondary, we see the expected shift towards $q_{\rm b} = 1$ in the CBD-evolved populations, most notably in those with higher accretion rates. Despite the minor shift in total mass, the mass ratio increase can lead to a significant increase in chirp masses, allowing more MBHBs to be detectable in LISA. 
In the rightmost panel we show the orbital eccentricity distributions. As all populations are initialized at a low orbital eccentricity $e_{\rm b,0} = 0.01$, the \texttt{nCBD} binaries enter the LISA band with undetectable residual eccentricities. However, even at low accretion rates (1\% Eddington; pale green line), CBD-driven eccentricity evolution is efficient enough to shift the distribution by over an order of magnitude. At 100\% Eddington, the residual eccentricities in the LISA band are near $10^{-3}$, close to the regime where detection of orbital eccentricity is expected \citep[e.g.,][]{Garg2023}. We point out that residual eccentricities are higher in lower mass binary systems (see Figure \ref{fig:freq_sepa_eb_mass_binned}), and plan to investigate the impact of CBD physics on IMBHs in future studies.

\begin{figure*}
    \centering
\includegraphics[width=1.0\textwidth]{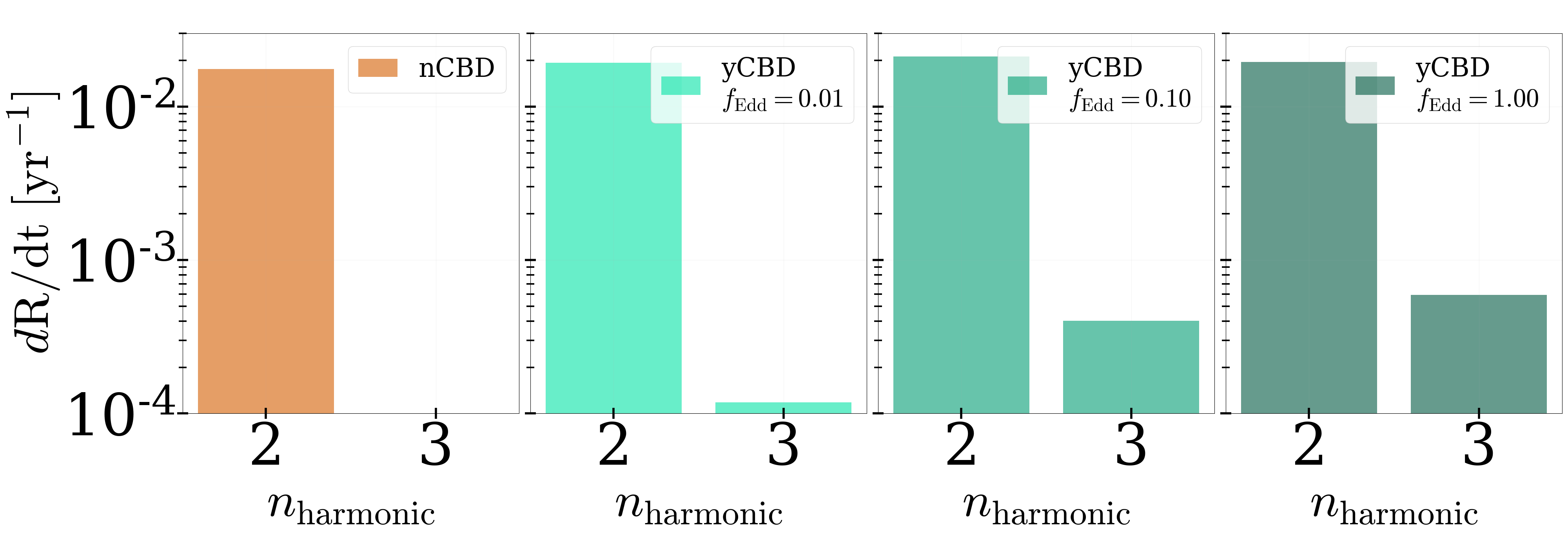}
    \caption{The rate of detected harmonics per year in the LISA band. We show models without CBDs (left), and with CBDs and increasing accretion rates (1\%, 10\% and 100\% Eddington going left to right). The detection rate of n=3 harmonics increases with the total binary accretion rate.}
\label{fig:lisa_detection_rates_fedd_study}
\end{figure*}

In Figure \ref{fig:lisa_detection_rates_fedd_study} we show the rate of mergers detected in our binary sample per harmonic in which they were detected. When comparing the \texttt{yCBD}/\texttt{nCBD} (green and orange respectively), we find that the presence of a CBD is associated with a higher rate of detected n=3 harmonics. We find that CBD driven evolution causes detection of the n=3 harmonic for a few percent of the detected n=2 binaries, especially if CBD accretion rates are high.
We note that all binaries that were detected in the n=3 harmonic were also detected in the n=2 harmonic, i.e. there are no MBHBs in our sample that are only detectable in LISA due to the excitation of an eccentric harmonic. However, this may change in lower mass binaries, where higher residual eccentricities may be expected (see Figure \ref{fig:eb_vs_mb_PTA_LSST_mb_eb}).

We note that our analysis serves to demonstrate that many binaries that co-evolve with CBDs emit enough GW power in their higher harmonics to be detectable. This is also a very conservative demonstration as a templated search will be much more sensitive to detecting eccentricity than simply searching for particular $n>2$ harmonics.

\begin{figure*}
    \centering
\includegraphics[width=1.0\textwidth]{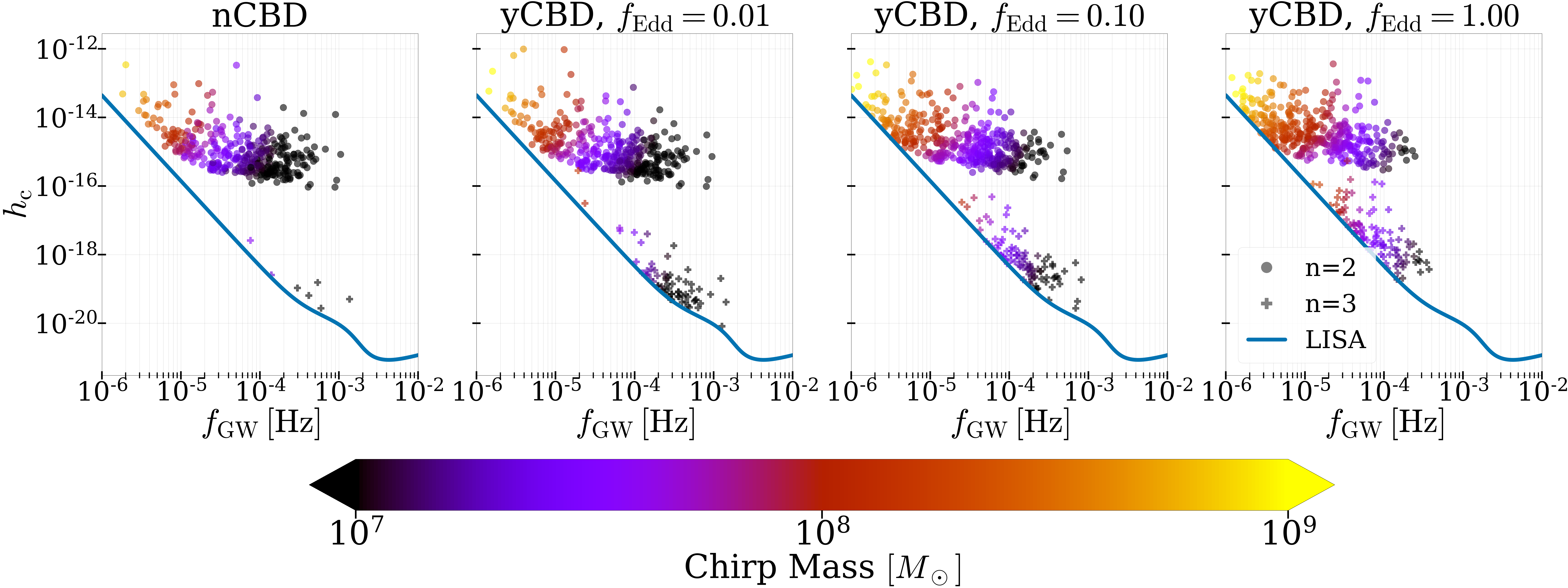}
    \caption{Detectable binaries in the LISA band, showing colour-coded masses of binaries detected in n=2 (circles) and n=3 (stars) harmonics. The higher the accretion rate, the more binaries are detected in eccentric harmonics.}
\label{fig:lisa_sensitivity}
\end{figure*}

% \subsection{Eddington limited accretion}
% Since accretion from a CBD preferentially occurs onto the secondary (e.g. \citep{Farris2014, Duffell2020, Siwek2023a}), $\dot{M}_2$ can exceed the Eddington limit. This is especially likely in binaries with very small mass ratios. Here we investigate whether Eddington limited accretion impacts the statistical properties of the MBHB populations we evolve. In Figure \ref{fig:qb_eddington} we show the mass ratio distributions of binaries in our light cone. We contrast MBHB populations where the accretion rate onto the secondary is unbound ($f_{\rm Edd, 2} = \infty$; green line), or limited by up to a factor of $1$ and $10$ (blue and orange lines, respectively). We find that the effect is negligible: super-Eddington accretion onto the secondary has only a small effect on the statistical properties of the observed population. We note that in Figure \ref{fig:qb_eddington}, we have assumed constant accretion at the Eddington limit of each binary. However, throughout our analysis, we assume a constant accretion rate at $10\%$ Eddington. At lower total accretion rates such as the fiducial $10\%$ Eddington, Figure \ref{fig:qb_eddington} would show no differences between populations evolved with different upper limits on the secondary accretion rate.

\section{Summary and Discussion}
\label{sec:discussion}
We have examined the impact of CBD driven evolution and preferential accretion on the population statistics of potentially observable MBHBs, the GWB and LISA binaries. Our analysis is based on a sample of MBHBs derived from galaxy mergers in the \texttt{Illustris} simulation, and semi-analytically evolved using \texttt{holodeck} \citep{Kelley2017MNRAS}. We have performed parameter studies testing several initial conditions, with initial eccentricities $e_{\rm b,0}$ either randomly drawn or uniformly set to the following values; $e_{\rm b,0} = [0.0001, 0.001, 0.01, 0.9]$ (see Figures \ref{fig:init_eb0_gw_freq_eb}, \ref{fig:hmap_LSST_PTA}). Our fiducial case forms MBHBs at near circular binaries with $e_{\rm b,0} = 0.01$. As the impact of CBD-driven orbital evolution is scaled by the total binary accretion rate \citep[for further details we refer to the orbital evolution models presented in][]{Siwek2023a,Siwek2023b}, we tested various levels Eddington rates at $f_{\rm Edd} = [0.01, 0.10, 1.0]$ (see Figures \ref{fig:eb_qb_dist}, \ref{fig:hmap_LSST_PTA}, \ref{fig:gwb_fedd}). Our fiducial case is $f_{\rm Edd} = 0.10$. Finally, the orbital evolution models our population study is based on contain semi-major axis evolution rates of binaries with a range of mass ratios and eccentricities. We therefore \textbf{self consistently allow for binary hardening and softening}, and present the results of self-consistent binary softening on the GWB (see Figure \ref{fig:gwb_softening}). Our analysis compares MBHBs evolved semi-analytically with large-scale galactic processes, CBDs and GWs to a population of MBHBs evolved without any CBD contribution. Using this parameter study we have identified signatures of CBD-driven binary evolution for upcoming multi-messenger observations of MBHBs.
Our key results are, 
\begin{enumerate}
    \item CBD dynamics dominate the eccentricity evolution of sub-parsec MBHBs, which evolve up to orbital eccentricities $e_{\rm b} \sim 0.5 $ as they enter the PTA band. 
    \item When CBD physics is applied, the eccentricities of MBHBs in the observable bands are independent of their highly uncertain initial eccentricity distribution at formation, erasing any memory of it.
    \item CBD-driven semi-major axis evolution has a minor effect on the GWB, counter-intuitively leading to a small boost in the GWB amplitude due to prolonged accretion.
    \item The recently discovered $e_{\rm b}$ versus $q_{\rm b}$ relationship induced by CBD dynamics \citep{Siwek2023b} leaves an unambiguous signature in detectable MBHB populations.
    \item MBHBs in the LISA band retain eccentricity from the CBD-driven evolution, boosting the median eccentricity of the population by up to 2 orders of magnitude.
    \item Increased orbital eccentricities enhance the likelihood that eccentricities are detected in the LISA band. 
\end{enumerate}
In the following, we put our findings in the context of the relevant literature. We outline any caveats the reader may wish to consider when interpreting our results.

\subsection{Comparison with the Literature}
We have shown that CBD dynamics is a non-negligible contributor to MBHB eccentricity evolution. Our results are based on the first comprehensive MBHB population synthesis model that uses numerically derived CBD-driven orbital evolution rates of binaries with arbitrary mass ratios and eccentricities \citep{Siwek2023a, Siwek2023b}. Prior studies have investigated the eccentricity evolution of LISA and PTA binaries without generalized CBD-driven evolution \citep[e.g.,][]{Sesana2010, Sayeb2021}, and we outline the main differences and agreement with this work.

\cite{Sesana2010} have conducted a detailed population analysis of MBHBs with varying mass ratios and eccentricities, evolved semi-analytically with stellar scattering models \citep{Sesana2008} and GW emission \citep{PetersMathews1963}. Similar to our work, \cite{Sesana2010} find that low mass binaries retain the highest eccentricities at observable frequencies (see Figure \ref{fig:eb_vs_mb_PTA_LSST_mb_eb}). 
Conversely, they find that low mass ratio binaries enter observable bands at higher eccentricities, while we find the opposite (Figure \ref{fig:hmap_LSST_PTA}). Our results include the influence of CBD-driven evolution on the orbital eccentricity, and thus deviate from their results. While we use the same stellar scattering models that favour eccentricity pumping in low mass ratio systems, we find that CBD-driven evolution is more significant. CBDs evolve binaries towards eccentricities that increase with mass ratio, and we find that this signature mechanism of CBD physics remains visible in PTA and LSST bands (see Figure \ref{fig:hmap_LSST_PTA}). We further find that very few low mass ratio systems with $q_{\rm b} \lesssim 0.6$ remain by the time binaries have evolved towards the detectable bands. We explain that this is due to CBD-driven preferential accretion on the secondary. 
\cite{Sesana2010}, on the other hand, do not include CBD-driven accretion to evolve binary mass ratios. They also cover a larger parameter space with mass ratios as low as $q_{\rm b} = 1/729$, and as a result their sample contains a larger number of very low mass ratio binary systems. As low mass ratio binaries evolve towards the highest eccentricities due to stellar scattering, they predict higher residual eccentricities in the LISA band than our simulations (see Figure \ref{fig:lisa_population_fedd_study}). 

Commenting on the dearth of low mass ratio systems in out sample compared to \cite{Sesana2010}, we point out that CBD accretion may not always lead to an increase in mass ratio. Instead, a turnover in the preferential accretion ratio \citep{Farris2014} could lead to a bi-modal mass ratio distribution in MBHBs \citep[as seen in][]{Siwek2020}. Further numerical studies to are needed to investigate this turnover, as preferential accretion models have only been tested numerically down to $q_{\rm b} \sim 0.01$ \citep[e.g.,][]{Farris2014,Duffell2020}. While these simulations hint at a potential turnover in the preferential accretion ratios, and could thus lead to a bi-modal mass ratio distribution, further studies with a focus on low mass ratio binaries are needed to confirm this. 

\cite{Sesana2010} find that the properties of MBHBs in observable frequency bands depend on the initial orbital eccentricities in their population. However, a main result of our work is that the observed eccentricities in MBHB populations do not depend on initial conditions, as a result of CBDs. CBD-driven evolution erases the initial eccentricity distribution in MBHBs, and thus the initial eccentricity distribution is irrelevant for the properties of observable MBHB populations. This both disagrees with \cite{Sesana2010}, and vastly simplifies parameter studies of MBHB populations, if CBD-driven evolution is indeed a common pathway in sub-parsec MBHBs. 

\cite{Roedig2011} conducted a numerical study which suggested a steady-state, or ``limiting" eccentricity in MBHBs due to CBD-driven evolution in the range $0.6-0.8$. They evolved binaries from the limiting eccentricity with GW-driven orbital evolution rates, and as a result estimates the eccentricity of post-CBD binaries in the LISA band to be $10^{-3} - 10^{-2}$ in low-mass binaries with $M_{\rm b} \lesssim 10^{6}\, M_{\odot}$. Given the slightly higher steady-state eccentricity and low binary mass, their results appear to be in line with our estimates (see Figure \ref{fig:eb_vs_mb_PTA_LSST_mb_eb}). 

\cite{Bortolas2021} investigated the effect of stars and CBD-driven orbital evolution on the GW detection prospects of circular, equal-mass MBHBs. Although they do not account for binary eccentricities, we can draw some parallels to our work. Similar to our findings, they conclude that allowing for CBD-driven orbital expansion has no meaningful effect on the merger timescales and thus the prospects of GW detections, which is also reflected in the GWB calculation shown in our Figure \ref{fig:gwb_softening}. However, they predict that high accretion rates may lead to a larger number of low-mass MBHBs detectable in the LISA band. They attribute this to the overall mass growth, while our analysis predicts an increase in eccentric merger detections due to accretion (see Figures \ref{fig:lisa_detection_rates_fedd_study} and \ref{fig:lisa_sensitivity}).

% Detectability: \cite{Garg2023} show that eccentricity is detectable down to $\sim 10^{-2.75}$ in low mass binary systems (masses around $10^{5}\, M_{\odot}$). Our model suggests that residual eccentricities will be higher in lower mass systems, and if this trend continues for binaries below our lower limit $M_{\rm MBHB} \lesssim 10^{-6} M_{\odot}$, significant eccentricities could be picked up by LISA (compare also with similar results in \cite{Zrake2021}).
 
\subsection{Caveats}
We have made several simplifications and assumptions in our simulations, which should be taken into account when interpreting our results. 

Our models assume that the CBD is aligned with the binary angular momentum vector, leading to prograde rotation. However, gas inflow can occur from arbitrary angles (chaotic accretion), and may thus result in inclined or retrograde accretion flows. Recently, \cite{TiedeDOrazio2023_arxiv} showed that equal mass ratio binaries in circumbinary disks can evolve to much higher eccentricities than in the unbound case. This may lead to much higher MBHB eccentricities, and residual eccentricities in the PTA, LSST and LISA bands may increase accordingly. 

Our models do not currently account for the case of MBH triples forming as a result of repeated galaxy mergers \citep[e.g.,][]{Bonetti2018a, Bonetti2018b, Sayeb2024MNRAS.527.7424S}. If triples form frequently, the merger timescales as well as eccentricity evolution of the inner binary can be affected significantly. Such binary systems may show up as outliers in population studies, with above-average residual eccentricities close to merger, but further numerical studies are needed to better understand the interplay between CBD dynamics and three-body interactions. 

We do not currently account for luminosities or AGN fuelling times of MBHBs in our population simulations to calculate the rates at which these systems may be detectable with LSST. Optical luminosities from circumbinary disks are uncertain, and depend on the radiative processes in the inner and outer disks, which are poorly understood and depend on hydrodynamic modeling of the accretion disk \citep[e.g.,][]{Krauth2023MNRAS.526.5441K}. Generally however, higher mass binaries may be more luminous and potentially more easily detected. In future surveys, this could lead to a systematic shift in the eccentricity distribution of MBHBs as high mass MBHBs are expected to be on less eccentric orbits (see Figure \ref{fig:eb_vs_mb_PTA_LSST_mb_eb}). Related to this, our MBHB sample is based on the \texttt{Illustris} simulation, and as a result has a cut on low mass MBHs. Including more low mass MBHs in population studies would likely increase the median eccentricities of MBHBs in observable bands significantly, and lead to more stringent constraints of CBD-driven residual eccentricities in LISA mergers.

\section{Data Availability}
The data underlying this article will be shared on reasonable request to the corresponding author.

%{\it\color{lightgray} \lipsum[10-17]} %%%%%%%%% FILLER TEXT

%%%%%%%%%%%%%%%%%%%%%%%%%%%%%%%
%%%%%%%% BIBLIOGRAPHY %%%%%%%%%
%\begingroup
%\let\clearpage\relax
\twocolumn

\bibliographystyle{mnras}
\bibliography{mybib} % if your bibtex file is called mybib.bib

\newcommand{\noop}[1]{}
\begin{thebibliography}{}
\makeatletter
\relax
\def\mn@urlcharsother{\let\do\@makeother \do\$\do\&\do\#\do\^\do\_\do\%\do\~}
\def\mn@doi{\begingroup\mn@urlcharsother \@ifnextchar [ {\mn@doi@}
  {\mn@doi@[]}}
\def\mn@doi@[#1]#2{\def\@tempa{#1}\ifx\@tempa\@empty \href
  {http://dx.doi.org/#2} {doi:#2}\else \href {http://dx.doi.org/#2} {#1}\fi
  \endgroup}
\def\mn@eprint#1#2{\mn@eprint@#1:#2::\@nil}
\def\mn@eprint@arXiv#1{\href {http://arxiv.org/abs/#1} {{\tt arXiv:#1}}}
\def\mn@eprint@dblp#1{\href {http://dblp.uni-trier.de/rec/bibtex/#1.xml}
  {dblp:#1}}
\def\mn@eprint@#1:#2:#3:#4\@nil{\def\@tempa {#1}\def\@tempb {#2}\def\@tempc
  {#3}\ifx \@tempc \@empty \let \@tempc \@tempb \let \@tempb \@tempa \fi \ifx
  \@tempb \@empty \def\@tempb {arXiv}\fi \@ifundefined
  {mn@eprint@\@tempb}{\@tempb:\@tempc}{\expandafter \expandafter \csname
  mn@eprint@\@tempb\endcsname \expandafter{\@tempc}}}

\bibitem[\protect\citeauthoryear{{Agazie} et~al.,}{{Agazie}
  et~al.}{2023a}]{NANOGravDetection2023}
{Agazie} G.,  et~al., 2023a, \mn@doi [\apjl] {10.3847/2041-8213/acdac6}, \href
  {https://ui.adsabs.harvard.edu/abs/2023ApJ...951L...8A} {951, L8}

\bibitem[\protect\citeauthoryear{{Agazie} et~al.,}{{Agazie}
  et~al.}{2023b}]{NANOGrav2023_CW}
{Agazie} G.,  et~al., 2023b, \mn@doi [\apjl] {10.3847/2041-8213/ace18a}, \href
  {https://ui.adsabs.harvard.edu/abs/2023ApJ...951L..50A} {951, L50}

\bibitem[\protect\citeauthoryear{{Agazie} et~al.,}{{Agazie}
  et~al.}{2023c}]{NANOGravMBHBs2023}
{Agazie} G.,  et~al., 2023c, \mn@doi [\apjl] {10.3847/2041-8213/ace18b}, \href
  {https://ui.adsabs.harvard.edu/abs/2023ApJ...952L..37A} {952, L37}

\bibitem[\protect\citeauthoryear{{Amaro-Seoane} et~al.,}{{Amaro-Seoane}
  et~al.}{2017}]{AmaroSeoane2017}
{Amaro-Seoane} P.,  et~al., 2017, arXiv e-prints, \href
  {https://ui.adsabs.harvard.edu/abs/2017arXiv170200786A} {p. arXiv:1702.00786}

\bibitem[\protect\citeauthoryear{Antoniadis et~al.,}{Antoniadis
  et~al.}{2023}]{EPTADetection2023}
Antoniadis J.,  et~al., 2023, The second data release from the European Pulsar
  Timing Array III. Search for gravitational wave signals,
  \mn@doi{10.48550/ARXIV.2306.16214}, \url {https://arxiv.org/abs/2306.16214}

\bibitem[\protect\citeauthoryear{Barnes \& Hernquist}{Barnes \&
  Hernquist}{1991}]{Barnes1991}
Barnes J.~E.,  Hernquist L.~E.,  1991, \mn@doi [The Astrophysical Journal]
  {10.1086/185978}, 370, L65

\bibitem[\protect\citeauthoryear{Barnes \& Hernquist}{Barnes \&
  Hernquist}{1996}]{Barnes1996}
Barnes J.~E.,  Hernquist L.,  1996, \mn@doi [The Astrophysical Journal]
  {10.1086/177957}, 471, 115

\bibitem[\protect\citeauthoryear{Begelman, Blandford  \& Rees}{Begelman
  et~al.}{1980}]{Begelman1980}
Begelman M.~C.,  Blandford R.~D.,   Rees M.~J.,  1980, \mn@doi [Nature]
  {10.1038/287307a0}, 287, 307

\bibitem[\protect\citeauthoryear{Bonetti, Sesana, Barausse  \& Haardt}{Bonetti
  et~al.}{2018a}]{Bonetti2018b}
Bonetti M.,  Sesana A.,  Barausse E.,   Haardt F.,  2018a, \mn@doi [Monthly
  Notices of the Royal Astronomical Society] {10.1093/mnras/sty874}, 477, 2599

\bibitem[\protect\citeauthoryear{{Bonetti}, {Haardt}, {Sesana}  \&
  {Barausse}}{{Bonetti} et~al.}{2018b}]{Bonetti2018a}
{Bonetti} M.,  {Haardt} F.,  {Sesana} A.,   {Barausse} E.,  2018b, \mn@doi
  [\mnras] {10.1093/mnras/sty896}, \href
  {https://ui.adsabs.harvard.edu/abs/2018MNRAS.477.3910B} {477, 3910}

\bibitem[\protect\citeauthoryear{{Bortolas}, {Franchini}, {Bonetti}  \&
  {Sesana}}{{Bortolas} et~al.}{2021}]{Bortolas2021}
{Bortolas} E.,  {Franchini} A.,  {Bonetti} M.,   {Sesana} A.,  2021, \mn@doi
  [\apjl] {10.3847/2041-8213/ac1c0c}, \href
  {https://ui.adsabs.harvard.edu/abs/2021ApJ...918L..15B} {918, L15}

\bibitem[\protect\citeauthoryear{{Chen} et~al.,}{{Chen}
  et~al.}{2022}]{Chen2022}
{Chen} N.,  et~al., 2022, \mn@doi [\mnras] {10.1093/mnras/stac1432}, \href
  {https://ui.adsabs.harvard.edu/abs/2022MNRAS.514.2220C} {514, 2220}

\bibitem[\protect\citeauthoryear{D'Orazio \& Duffell}{D'Orazio \&
  Duffell}{2021}]{DorazioDuffell2021}
D'Orazio D.~J.,  Duffell P.~C.,  2021, \mn@doi [The Astrophysical Journal]
  {10.3847/2041-8213/ac0621}, 914, L21

\bibitem[\protect\citeauthoryear{{Delgado} et~al.,}{{Delgado}
  et~al.}{2023}]{Delgado_MTNG2023MNRAS.523.5899D}
{Delgado} A.~M.,  et~al., 2023, \mn@doi [\mnras] {10.1093/mnras/stad1781},
  \href {https://ui.adsabs.harvard.edu/abs/2023MNRAS.523.5899D} {523, 5899}

\bibitem[\protect\citeauthoryear{{Dotti}, {Colpi}, {Haardt}  \&
  {Mayer}}{{Dotti} et~al.}{2007}]{Dotti2007}
{Dotti} M.,  {Colpi} M.,  {Haardt} F.,   {Mayer} L.,  2007, \mn@doi [\mnras]
  {10.1111/j.1365-2966.2007.12010.x}, \href
  {https://ui.adsabs.harvard.edu/abs/2007MNRAS.379..956D} {379, 956}

\bibitem[\protect\citeauthoryear{Duffell, D'Orazio, Derdzinski, Haiman,
  MacFadyen, Rosen  \& Zrake}{Duffell et~al.}{2020}]{Duffell2020}
Duffell P.~C.,  D'Orazio D.,  Derdzinski A.,  Haiman Z.,  MacFadyen A.,  Rosen
  A.~L.,   Zrake J.,  2020, \mn@doi [The Astrophysical Journal]
  {10.3847/1538-4357/abab95}, 901, 25

\bibitem[\protect\citeauthoryear{Enoki \& Nagashima}{Enoki \&
  Nagashima}{2007}]{Enoki2007}
Enoki M.,  Nagashima M.,  2007, \mn@doi [Progress of Theoretical Physics]
  {10.1143/ptp.117.241}, 117, 241

\bibitem[\protect\citeauthoryear{Farris, Duffell, MacFadyen  \& Haiman}{Farris
  et~al.}{2014}]{Farris2014}
Farris B.~D.,  Duffell P.,  MacFadyen A.~I.,   Haiman Z.,  2014, \mn@doi [The
  Astrophysical Journal] {10.1088/0004-637x/783/2/134}, 783, 134

\bibitem[\protect\citeauthoryear{{Foster} \& {Backer}}{{Foster} \&
  {Backer}}{1990}]{Foster1990}
{Foster} R.~S.,  {Backer} D.~C.,  1990, in Bulletin of the American
  Astronomical Society. p.~1341

\bibitem[\protect\citeauthoryear{Garg, Tiwari, Derdzinski, Baker, Marsat  \&
  Mayer}{Garg et~al.}{2023}]{Garg2023}
Garg M.,  Tiwari S.,  Derdzinski A.,  Baker J.,  Marsat S.,   Mayer L.,  2023,
  The minimum measurable eccentricity from gravitational waves of LISA massive
  black hole binaries (\mn@eprint {} {arXiv:2307.13367})

\bibitem[\protect\citeauthoryear{Genel et~al.,}{Genel et~al.}{2014}]{Genel2014}
Genel S.,  et~al., 2014, \mn@doi [Monthly Notices of the Royal Astronomical
  Society] {10.1093/mnras/stu1654}, 445, 175

\bibitem[\protect\citeauthoryear{{Goodman}}{{Goodman}}{2003}]{Goodman2003MNRAS.339..937G}
{Goodman} J.,  2003, \mn@doi [\mnras] {10.1046/j.1365-8711.2003.06241.x}, \href
  {https://ui.adsabs.harvard.edu/abs/2003MNRAS.339..937G} {339, 937}

\bibitem[\protect\citeauthoryear{Gualandris, Read, Dehnen  \&
  Bortolas}{Gualandris et~al.}{2016}]{Gualandris2016}
Gualandris A.,  Read J.~I.,  Dehnen W.,   Bortolas E.,  2016, \mn@doi [Monthly
  Notices of the Royal Astronomical Society] {10.1093/mnras/stw2528}, 464, 2301

\bibitem[\protect\citeauthoryear{{Gualandris}, {Khan}, {Bortolas}, {Bonetti},
  {Sesana}, {Berczik}  \& {Holley-Bockelmann}}{{Gualandris}
  et~al.}{2022}]{Gualandris2022}
{Gualandris} A.,  {Khan} F.~M.,  {Bortolas} E.,  {Bonetti} M.,  {Sesana} A.,
  {Berczik} P.,   {Holley-Bockelmann} K.,  2022, \mn@doi [\mnras]
  {10.1093/mnras/stac241}, \href
  {https://ui.adsabs.harvard.edu/abs/2022MNRAS.511.4753G} {511, 4753}

\bibitem[\protect\citeauthoryear{Haiman, Kocsis  \& Menou}{Haiman
  et~al.}{2009}]{Haiman2009}
Haiman Z.,  Kocsis B.,   Menou K.,  2009, \mn@doi [The Astrophysical Journal]
  {10.1088/0004-637x/700/2/1952}, 700, 1952

\bibitem[\protect\citeauthoryear{{Ivezi{\'c}} et~al.,}{{Ivezi{\'c}}
  et~al.}{2019}]{LSST2019}
{Ivezi{\'c}} {\v Z}.,  et~al., 2019, \mn@doi [\apj] {10.3847/1538-4357/ab042c},
  \href {http://adsabs.harvard.edu/abs/2019ApJ...873..111I} {873, 111}

\bibitem[\protect\citeauthoryear{{Katz} \& {Larson}}{{Katz} \&
  {Larson}}{2019}]{KatzLarson2019}
{Katz} M.~L.,  {Larson} S.~L.,  2019, \mn@doi [\mnras] {10.1093/mnras/sty3321},
  \href {https://ui.adsabs.harvard.edu/abs/2019MNRAS.483.3108K} {483, 3108}

\bibitem[\protect\citeauthoryear{{Katz}, {Kelley}, {Dosopoulou}, {Berry},
  {Blecha}  \& {Larson}}{{Katz} et~al.}{2020}]{Katz2020}
{Katz} M.~L.,  {Kelley} L.~Z.,  {Dosopoulou} F.,  {Berry} S.,  {Blecha} L.,
  {Larson} S.~L.,  2020, \mn@doi [\mnras] {10.1093/mnras/stz3102}, \href
  {https://ui.adsabs.harvard.edu/abs/2020MNRAS.491.2301K} {491, 2301}

\bibitem[\protect\citeauthoryear{Kelley, Blecha  \& Hernquist}{Kelley
  et~al.}{2016}]{Kelley2016}
Kelley L.~Z.,  Blecha L.,   Hernquist L.,  2016, \mn@doi [Monthly Notices of
  the Royal Astronomical Society] {10.1093/mnras/stw2452}, 464, 3131

\bibitem[\protect\citeauthoryear{{Kelley}, {Blecha}  \& {Hernquist}}{{Kelley}
  et~al.}{2017a}]{Kelley2017MNRAS}
{Kelley} L.~Z.,  {Blecha} L.,   {Hernquist} L.,  2017a, \mn@doi [\mnras]
  {10.1093/mnras/stw2452}, \href
  {https://ui.adsabs.harvard.edu/abs/2017MNRAS.464.3131K} {464, 3131}

\bibitem[\protect\citeauthoryear{Kelley, Blecha, Hernquist, Sesana  \&
  Taylor}{Kelley et~al.}{2017b}]{Kelley2017}
Kelley L.~Z.,  Blecha L.,  Hernquist L.,  Sesana A.,   Taylor S.~R.,  2017b,
  \mn@doi [Monthly Notices of the Royal Astronomical Society]
  {10.1093/mnras/stx1638}, 471, 4508

\bibitem[\protect\citeauthoryear{{Kelley}, {Blecha}, {Hernquist}, {Sesana}  \&
  {Taylor}}{{Kelley} et~al.}{2018}]{Kelley2018MNRAS.477..964K}
{Kelley} L.~Z.,  {Blecha} L.,  {Hernquist} L.,  {Sesana} A.,   {Taylor} S.~R.,
  2018, \mn@doi [\mnras] {10.1093/mnras/sty689}, \href
  {https://ui.adsabs.harvard.edu/abs/2018MNRAS.477..964K} {477, 964}

\bibitem[\protect\citeauthoryear{Kelley, Haiman, Sesana  \& Hernquist}{Kelley
  et~al.}{2019}]{Kelley2019}
Kelley L.~Z.,  Haiman Z.,  Sesana A.,   Hernquist L.,  2019, \mn@doi [Monthly
  Notices of the Royal Astronomical Society] {10.1093/mnras/stz150}, 485, 1579

\bibitem[\protect\citeauthoryear{Khan, Just  \& Merritt}{Khan
  et~al.}{2011}]{Khan2011}
Khan F.~M.,  Just A.,   Merritt D.,  2011, \mn@doi [The Astrophysical Journal]
  {10.1088/0004-637x/732/2/89}, 732, 89

\bibitem[\protect\citeauthoryear{Kocsis \& Sesana}{Kocsis \&
  Sesana}{2011}]{Kocsis2011}
Kocsis B.,  Sesana A.,  2011, \mn@doi [Monthly Notices of the Royal
  Astronomical Society] {10.1111/j.1365-2966.2010.17782.x}, 411, 1467

\bibitem[\protect\citeauthoryear{{Krauth}, {Davelaar}, {Haiman},
  {Westernacher-Schneider}, {Zrake}  \& {MacFadyen}}{{Krauth}
  et~al.}{2023}]{Krauth2023MNRAS.526.5441K}
{Krauth} L.~M.,  {Davelaar} J.,  {Haiman} Z.,  {Westernacher-Schneider} J.~R.,
  {Zrake} J.,   {MacFadyen} A.,  2023, \mn@doi [\mnras]
  {10.1093/mnras/stad3095}, \href
  {https://ui.adsabs.harvard.edu/abs/2023MNRAS.526.5441K} {526, 5441}

\bibitem[\protect\citeauthoryear{{Madigan} \& {Levin}}{{Madigan} \&
  {Levin}}{2012}]{MadiganLevin2012}
{Madigan} A.-M.,  {Levin} Y.,  2012, \mn@doi [\apj]
  {10.1088/0004-637X/754/1/42}, \href
  {https://ui.adsabs.harvard.edu/abs/2012ApJ...754...42M} {754, 42}

\bibitem[\protect\citeauthoryear{{Marconi}, {Risaliti}, {Gilli}, {Hunt},
  {Maiolino}  \& {Salvati}}{{Marconi}
  et~al.}{2004}]{Marconi2004MNRAS.351..169M}
{Marconi} A.,  {Risaliti} G.,  {Gilli} R.,  {Hunt} L.~K.,  {Maiolino} R.,
  {Salvati} M.,  2004, \mn@doi [\mnras] {10.1111/j.1365-2966.2004.07765.x},
  \href {https://ui.adsabs.harvard.edu/abs/2004MNRAS.351..169M} {351, 169}

\bibitem[\protect\citeauthoryear{{Marinacci} et~al.,}{{Marinacci}
  et~al.}{2018}]{Marinacci_TNG2018MNRAS.480.5113M}
{Marinacci} F.,  et~al., 2018, \mn@doi [\mnras] {10.1093/mnras/sty2206}, \href
  {https://ui.adsabs.harvard.edu/abs/2018MNRAS.480.5113M} {480, 5113}

\bibitem[\protect\citeauthoryear{Miranda, Mu{\~{n}}oz  \& Lai}{Miranda
  et~al.}{2016}]{Miranda2016}
Miranda R.,  Mu{\~{n}}oz D.~J.,   Lai D.,  2016, \mn@doi [Monthly Notices of
  the Royal Astronomical Society] {10.1093/mnras/stw3189}, 466, 1170

\bibitem[\protect\citeauthoryear{{Mirza}, {Tahir}, {Khan}, {Holley-Bockelmann},
  {Baig}, {Berczik}  \& {Chishtie}}{{Mirza} et~al.}{2017}]{Mirza2017}
{Mirza} M.~A.,  {Tahir} A.,  {Khan} F.~M.,  {Holley-Bockelmann} H.,  {Baig}
  A.~M.,  {Berczik} P.,   {Chishtie} F.,  2017, \mn@doi [\mnras]
  {10.1093/mnras/stx1248}, \href
  {https://ui.adsabs.harvard.edu/abs/2017MNRAS.470..940M} {470, 940}

\bibitem[\protect\citeauthoryear{Moody, Shi  \& Stone}{Moody
  et~al.}{2019}]{Moody2019}
Moody M. S.~L.,  Shi J.-M.,   Stone J.~M.,  2019, \mn@doi [The Astrophysical
  Journal] {10.3847/1538-4357/ab09ee}, 875, 66

\bibitem[\protect\citeauthoryear{{Mu{\~n}oz}, {Lai}, {Kratter}  \&
  {Miranda}}{{Mu{\~n}oz} et~al.}{2020}]{Munoz2020}
{Mu{\~n}oz} D.~J.,  {Lai} D.,  {Kratter} K.,   {Miranda} R.,  2020, \mn@doi
  [\apj] {10.3847/1538-4357/ab5d33}, \href
  {https://ui.adsabs.harvard.edu/abs/2020ApJ...889..114M} {889, 114}

\bibitem[\protect\citeauthoryear{Mu{\~{n}}oz, Miranda  \& Lai}{Mu{\~{n}}oz
  et~al.}{2019}]{Munoz2019}
Mu{\~{n}}oz D.~J.,  Miranda R.,   Lai D.,  2019, \mn@doi [The Astrophysical
  Journal] {10.3847/1538-4357/aaf867}, 871, 84

\bibitem[\protect\citeauthoryear{{Naiman} et~al.,}{{Naiman}
  et~al.}{2018}]{Naiman_TNG2018MNRAS.477.1206N}
{Naiman} J.~P.,  et~al., 2018, \mn@doi [\mnras] {10.1093/mnras/sty618}, \href
  {https://ui.adsabs.harvard.edu/abs/2018MNRAS.477.1206N} {477, 1206}

\bibitem[\protect\citeauthoryear{{Nelson} et~al.,}{{Nelson}
  et~al.}{2018}]{Nelson_TNG2018MNRAS.475..624N}
{Nelson} D.,  et~al., 2018, \mn@doi [\mnras] {10.1093/mnras/stx3040}, \href
  {https://ui.adsabs.harvard.edu/abs/2018MNRAS.475..624N} {475, 624}

\bibitem[\protect\citeauthoryear{Peters}{Peters}{1964}]{Peters1964}
Peters P.~C.,  1964, \mn@doi [Physical Review] {10.1103/physrev.136.b1224},
  136, B1224

\bibitem[\protect\citeauthoryear{Peters \& Mathews}{Peters \&
  Mathews}{1963}]{PetersMathews1963}
Peters P.~C.,  Mathews J.,  1963, \mn@doi [Phys. Rev.]
  {10.1103/PhysRev.131.435}, 131, 435

\bibitem[\protect\citeauthoryear{{Pillepich} et~al.,}{{Pillepich}
  et~al.}{2018}]{Pillepich_TNG2018MNRAS.475..648P}
{Pillepich} A.,  et~al., 2018, \mn@doi [\mnras] {10.1093/mnras/stx3112}, \href
  {https://ui.adsabs.harvard.edu/abs/2018MNRAS.475..648P} {475, 648}

\bibitem[\protect\citeauthoryear{{Porter} \& {Sesana}}{{Porter} \&
  {Sesana}}{2010}]{Porter_Sesana2010}
{Porter} E.~K.,  {Sesana} A.,  2010, \mn@doi [arXiv e-prints]
  {10.48550/arXiv.1005.5296}, \href
  {https://ui.adsabs.harvard.edu/abs/2010arXiv1005.5296P} {p. arXiv:1005.5296}

\bibitem[\protect\citeauthoryear{Quinlan \& Hernquist}{Quinlan \&
  Hernquist}{1997}]{Quinlan1997}
Quinlan G.~D.,  Hernquist L.,  1997, \mn@doi [New Astronomy]
  {10.1016/s1384-1076(97)00039-0}, 2, 533

\bibitem[\protect\citeauthoryear{Rawlings, Mannerkoski, Johansson  \&
  Naab}{Rawlings et~al.}{2023}]{Rawlings2023}
Rawlings A.,  Mannerkoski M.,  Johansson P.~H.,   Naab T.,  2023, Reviving
  stochasticity: uncertainty in SMBH binary eccentricity is unavoidable
  (\mn@eprint {} {arXiv:2307.08756})

\bibitem[\protect\citeauthoryear{Reardon et~al.,}{Reardon
  et~al.}{2023}]{PPTADetection2023}
Reardon D.~J.,  et~al., 2023, \mn@doi [The Astrophysical Journal Letters]
  {10.3847/2041-8213/acdd02}, 951, L6

\bibitem[\protect\citeauthoryear{Roedig, Dotti, Sesana, Cuadra  \&
  Colpi}{Roedig et~al.}{2011}]{Roedig2011}
Roedig C.,  Dotti M.,  Sesana A.,  Cuadra J.,   Colpi M.,  2011, \mn@doi
  [Monthly Notices of the Royal Astronomical Society]
  {10.1111/j.1365-2966.2011.18927.x}, 415, 3033

\bibitem[\protect\citeauthoryear{{Sayeb}, {Blecha}, {Kelley}, {Gerosa},
  {Kesden}  \& {Thomas}}{{Sayeb} et~al.}{2021}]{Sayeb2021}
{Sayeb} M.,  {Blecha} L.,  {Kelley} L.~Z.,  {Gerosa} D.,  {Kesden} M.,
  {Thomas} J.,  2021, \mn@doi [\mnras] {10.1093/mnras/staa3826}, \href
  {https://ui.adsabs.harvard.edu/abs/2021MNRAS.501.2531S} {501, 2531}

\bibitem[\protect\citeauthoryear{{Sayeb}, {Blecha}  \& {Kelley}}{{Sayeb}
  et~al.}{2024}]{Sayeb2024MNRAS.527.7424S}
{Sayeb} M.,  {Blecha} L.,   {Kelley} L.~Z.,  2024, \mn@doi [\mnras]
  {10.1093/mnras/stad3637}, \href
  {https://ui.adsabs.harvard.edu/abs/2024MNRAS.527.7424S} {527, 7424}

\bibitem[\protect\citeauthoryear{{Sesana}}{{Sesana}}{2010}]{Sesana2010}
{Sesana} A.,  2010, \mn@doi [\apj] {10.1088/0004-637X/719/1/851}, \href
  {https://ui.adsabs.harvard.edu/abs/2010ApJ...719..851S} {719, 851}

\bibitem[\protect\citeauthoryear{{Sesana}}{{Sesana}}{2013}]{Sesana2013CQGra..30v4014S}
{Sesana} A.,  2013, \mn@doi [Classical and Quantum Gravity]
  {10.1088/0264-9381/30/22/224014}, \href
  {https://ui.adsabs.harvard.edu/abs/2013CQGra..30v4014S} {30, 224014}

\bibitem[\protect\citeauthoryear{Sesana, Haardt  \& Madau}{Sesana
  et~al.}{2006}]{Sesana2006}
Sesana A.,  Haardt F.,   Madau P.,  2006, \mn@doi [The Astrophysical Journal]
  {10.1086/507596}, 651, 392

\bibitem[\protect\citeauthoryear{Sesana, Vecchio  \& Colacino}{Sesana
  et~al.}{2008}]{Sesana2008}
Sesana A.,  Vecchio A.,   Colacino C.~N.,  2008, \mn@doi [Monthly Notices of
  the Royal Astronomical Society] {10.1111/j.1365-2966.2008.13682.x}, 390, 192

\bibitem[\protect\citeauthoryear{{Sesana}, {Gualandris}  \& {Dotti}}{{Sesana}
  et~al.}{2011}]{Sesana2011}
{Sesana} A.,  {Gualandris} A.,   {Dotti} M.,  2011, \mn@doi [\mnras]
  {10.1111/j.1745-3933.2011.01073.x}, \href
  {https://ui.adsabs.harvard.edu/abs/2011MNRAS.415L..35S} {415, L35}

\bibitem[\protect\citeauthoryear{{Siwek}, {Kelley}  \& {Hernquist}}{{Siwek}
  et~al.}{2020}]{Siwek2020}
{Siwek} M.~S.,  {Kelley} L.~Z.,   {Hernquist} L.,  2020, \mn@doi [\mnras]
  {10.1093/mnras/staa2361}, \href
  {https://ui.adsabs.harvard.edu/abs/2020MNRAS.498..537S} {498, 537}

\bibitem[\protect\citeauthoryear{{Siwek}, {Weinberger}, {Mu{\~n}oz}  \&
  {Hernquist}}{{Siwek} et~al.}{2023a}]{Siwek2023a}
{Siwek} M.,  {Weinberger} R.,  {Mu{\~n}oz} D.~J.,   {Hernquist} L.,  2023a,
  \mn@doi [\mnras] {10.1093/mnras/stac3263}, \href
  {https://ui.adsabs.harvard.edu/abs/2023MNRAS.518.5059S} {518, 5059}

\bibitem[\protect\citeauthoryear{{Siwek}, {Weinberger}  \& {Hernquist}}{{Siwek}
  et~al.}{2023b}]{Siwek2023b}
{Siwek} M.,  {Weinberger} R.,   {Hernquist} L.,  2023b, \mn@doi [\mnras]
  {10.1093/mnras/stad1131}, \href
  {https://ui.adsabs.harvard.edu/abs/2023MNRAS.522.2707S} {522, 2707}

\bibitem[\protect\citeauthoryear{{Springel} et~al.,}{{Springel}
  et~al.}{2018}]{Springel_TNG2018MNRAS.475..676S}
{Springel} V.,  et~al., 2018, \mn@doi [\mnras] {10.1093/mnras/stx3304}, \href
  {https://ui.adsabs.harvard.edu/abs/2018MNRAS.475..676S} {475, 676}

\bibitem[\protect\citeauthoryear{Tiede \& D'Orazio}{Tiede \&
  D'Orazio}{2023}]{TiedeDOrazio2023_arxiv}
Tiede C.,  D'Orazio D.~J.,  2023, Eccentric Binaries in Retrograde Disks,
  \mn@doi{10.48550/ARXIV.2307.03775}, \url {https://arxiv.org/abs/2307.03775}

\bibitem[\protect\citeauthoryear{Tiede, Zrake, MacFadyen  \& Haiman}{Tiede
  et~al.}{2020}]{Tiede2020}
Tiede C.,  Zrake J.,  MacFadyen A.,   Haiman Z.,  2020, \mn@doi [The
  Astrophysical Journal] {10.3847/1538-4357/aba432}, 900, 43

\bibitem[\protect\citeauthoryear{Vogelsberger et~al.,}{Vogelsberger
  et~al.}{2014a}]{Vogelsberger2014b}
Vogelsberger M.,  et~al., 2014a, \mn@doi [Monthly Notices of the Royal
  Astronomical Society] {10.1093/mnras/stu1536}, 444, 1518

\bibitem[\protect\citeauthoryear{Vogelsberger et~al.,}{Vogelsberger
  et~al.}{2014b}]{Vogelsberger2014a}
Vogelsberger M.,  et~al., 2014b, \mn@doi [Nature] {10.1038/nature13316}, 509,
  177

\bibitem[\protect\citeauthoryear{{Xin} \& {Haiman}}{{Xin} \&
  {Haiman}}{2021}]{Xin2021}
{Xin} C.,  {Haiman} Z.,  2021, \mn@doi [\mnras] {10.1093/mnras/stab1856}, \href
  {https://ui.adsabs.harvard.edu/abs/2021MNRAS.506.2408X} {506, 2408}

\bibitem[\protect\citeauthoryear{Xu et~al.,}{Xu
  et~al.}{2023}]{CPTADetection2023}
Xu H.,  et~al., 2023, \mn@doi [Research in Astronomy and Astrophysics]
  {10.1088/1674-4527/acdfa5}, 23, 075024

\bibitem[\protect\citeauthoryear{Yu}{Yu}{2002}]{Yu2002}
Yu Q.,  2002, \mn@doi [Monthly Notices of the Royal Astronomical Society]
  {10.1046/j.1365-8711.2002.05242.x}, 331, 935

\bibitem[\protect\citeauthoryear{{Yu} \& {Tremaine}}{{Yu} \&
  {Tremaine}}{2002}]{YuTremaine2002}
{Yu} Q.,  {Tremaine} S.,  2002, \mn@doi [\mnras]
  {10.1046/j.1365-8711.2002.05532.x}, \href
  {https://ui.adsabs.harvard.edu/abs/2002MNRAS.335..965Y} {335, 965}

\bibitem[\protect\citeauthoryear{{Zrake}, {Tiede}, {MacFadyen}  \&
  {Haiman}}{{Zrake} et~al.}{2021}]{Zrake2021}
{Zrake} J.,  {Tiede} C.,  {MacFadyen} A.,   {Haiman} Z.,  2021, \mn@doi [\apjl]
  {10.3847/2041-8213/abdd1c}, \href
  {https://ui.adsabs.harvard.edu/abs/2021ApJ...909L..13Z} {909, L13}

\makeatother
\end{thebibliography}
%\endgroup

%\label{lastpage}
\end{document}